# Spark Plasma Sintering for high-speed diffusion welding of the ultrafine grained near-α Ti-5Al-2V alloy with high strength and corrosion resistance for nuclear engineering

Vladimir Chuvil'deev[1], Aleksey Nokhrin[1*], Vladimir Kopylov[1,2], Maksim Boldin[1], Mikhail Vostokov[1], Mikhail Gryaznov[1], Nataliya Tabachkova[3], and Petr Tryaev[4]

[1] Lobachevsky State University of Nizhny Novgorod, 23 Gagarin Ave., Nizhniy Novgorod 603950 Russian Federation

[2] Physics and Technology Institute, National Academy of Sciences of Belarus, 10 Kuprevich St., Minsk 220141 Belarus

[3] National University of Science and Technology "MISIS", 4 Leninskiy Ave., Moscow 119049 Russian Federation

[4] Afrikantov OKBM JSC, Russian Nuclear Corporation "ROSATOM", 15 Burnakovsky Dr., Nizhnii Novgorod 15603074 Russian Federation

## Abstract

The paper demonstrates the prospects of Spark Plasma Sintering (SPS) for the high-speed diffusion welding of the high-strength ultrafine-grained (UFG) near-α Ti-5Al-2V alloy. The effect of increased diffusion welding intensity in the UFG Ti alloys is discussed also. The welds of the UFG near-α-Ti-5Al-2V alloy obtained by SPS are featured by high density, strength, and corrosion resistance. The rate of weld sealing in the UFG alloys has been shown to depend on the heating rate non-monotonously (with a pronounced maximum). At the stage of continuous heating and isothermic holding, the kinetics of the weld sealing was found to be determined by the exponential creep rate, the intensity of which in the coarse-grained (CG) alloys is limited by the diffusion rate in the crystal lattice whereas in the UFG alloys it is limited by the grain boundary diffusion rate.

---

* Corresponding author (nokhrin@nifti.unn.ru, ORCID: 0000-0002-0328-1923)
VC (chuvildeev@nifti.unn.ru, ORCID 0000-0002-5411-6557), VK (kopylov.ecap@gmail.com, ORCID 0000-0002-0725-9918), MB (boldin@nifti.unn.ru, ORCID 0000-0002-3156-8454), MV (m.vostokov96@yandex.ru, ORCID 0000-0001-9619-7150), MG (gryaznov@nifti.unn.ru, ORCID 0000-0001-7945-5419), NT (ntabachkova@gmail.com, ORCID 0000-0002-0169-5014), PT (ptryaev@okbm.nnov.ru, ORCID 0000-0001-6143-8223).

1. Introduction

Structural α- and near-α Ti alloys are used extensively in the production of the heat exchange equipment and other construction units for modern nuclear power plants (NPP) [1-6]. Besides, α- and near α-Ti alloys are used widely in petrochemistry, shipbuilding, in production of engineering structures operated at low temperatures, etc. [3-10], the requirements to the reliability and resources for which are getting stricter continuously. Therefore, the requirements to the mechanical and operational properties of the α- and near-α Ti alloys are growing.

To improve the strength of modern Ti alloys, the technologies based on the optimization of their composition and different types of thermo-mechanical processing modes are applied widely at present [3, 5-6]. One of the most promising ways to improve the mechanical properties of the Ti alloys is to form an ultrafine-grained (UFG) structure in these ones using different methods of severe plastic deformation (SPD), including the technology known as Equal Channel Angular Pressing (ECAP) [11-14], high pressure torsion [15-18], multiaxial forging [19, 20], rotary swaging [21, 22], etc. This approach ensures a unique combination of high strength, ductility, corrosion resistance, superplasticity, etc. This, in turn, opens the prospects to ensure a high performance and lifetime of the NPP heat exchange equipment.

One of the key challenges obstructing an extensive application of the UFG Ti alloys is the problem of welding. Traditional argon-arc welding [23, 24] or electron-beam welding [25, 26] accompanied by metal melting fail to preserve the UFG structure with high mechanical properties in the weld. Such solid-phase technologies as conventional diffusion welding [27, 28], friction stir welding [29-31] or superplastic forming [32, 33] require high temperatures and long holding times that leads to the growth of grains. Thus, all the unique properties of the UFG materials are lost. Note also that in the case of α- and near-α Ti alloys, the weld performance is impossible to improve through additional local heat treatment, which ensures the hardening of the weld as a result of the precipitation of the β-phase particles. In this case, the welds in the UFG α- and near-α Ti alloys have much poorer properties and therefore worsen the service life and safe operation of the entire steelwork.

Spark Plasma Sintering (SPS) offers better opportunities to obtain the high-strength welds since the idea underlying this innovative technology involves rapid heating (up to 2500 ºC/min) in vacuum or in an inert ambient. This effect is achieved by sending the high-power millisecond direct current (DC) pulses through the specimen along with applying an uniaxial pressure [34-38]. This technology produces the high-density structures in the UFG materials at lower optimal sintering temperatures [39-41]. High heating rates and faster diffusion at lower heating temperatures are crucial for restricting the grain growth and preserving the UFG structure in the material [34-41].

Note that obtaining a high-density weld as strong as the base metal is a scientific challenge. The high-density structure requires longer holding times at elevated temperatures, while the UFG structure can be preserved through restricting the grain growth by reducing the process time only. This problem can be solved by the diffusion mass transfer mechanism associated with the non-equilibrium grain boundaries during the rapid heating of the UFG metals.

This paper aims to explore the potential of SPS for the diffusion welding of the high-strength corrosion-resistant UFG α-Ti alloys used in nuclear engineering.

## 2. Materials and Methods

The study was focused on the Ti-4.73wt.%Al-1.88wt.%V alloy used in the production of the heat exchange equipment for modern NPPs. The chemical composition of the alloy is presented in Table 1. This alloy is one of near-α Ti alloys, the volume fraction of the β-phase in which does not exceed 5%. General chemical analysis of the alloy was carried out using FOUNDRY®-Master™ spectrum analyzer. The concentrations of oxygen, nitrogen, carbon, and hydrogen were measured by reductive melting using ELTRA® ON-900 ($O_2$, C, $N_2$) and ELTRA® OH-900 ($H_2$) analyzers.

The UFG structure in the alloy was formed by ECAP in the $B_c$ pressing mode (the number of the ECAP cycles $N = 4$, the ECAP temperature was 450 °C, the ECAP rate was 0.2 to 0.4 mm/s). ECAP was performed using Ficep® HF400L hydraulic press. Before pressing, the specimens of 14×14×140 mm³ in size were held in the operating channel of the ECAP setup for 10 min. The accuracy of temperature control during ECAP was ± 10 °C.

The diffusion welding of the specimens of 7×7×3.5 mm³ in size was performed using Dr. Sinter® SPS-625 setup (Fig. 1). The specimens for investigations were obtained by spark cutting into pieces of 14×14×140 mm³ in size. The difference in height between weld specimens did not exceed 0.02 mm. The surface roughness in the specimens was 40 to 60 µm. The scatter in the values of the surface roughness originated from the polishing with diamond polishing wax with different diamond grain dispersion. The decrease in the surface roughness leads to a slight decrease in the shrinkage $L$ without any significant change in the slope of the dependence of the shrinkage on the heating temperature $T$ $L(T)$ (Fig. 2). In further analysis of the experimental data, the effect of roughness was accounted for as an error when determining the diffusion welding activation energy (see Discussion section below).

The heating rate $V_h$ varied from 10 to 350 °C/min, the welding temperature $T$ ranged from 600 to 1140 °C, the applied pressure $P$ changed from 50 to 100 MPa. The degree of shrinkage $L_{eff}$ and shrinkage rate $S_{eff}$ were monitored by a dilatometer integrated into Dr. Sinter® SPS-625 setup. The accuracy of temperature control was ± 10 °C, the uncertainty of the shrinkage measurements was ± 0.01 mm. Welding was performed in vacuum (6 Pa).

To account for the contribution of the thermal expansion of the Dr. Sinter® SPS-625 sintering setup itself into the measured shrinkage $L_{eff}$, the measurements of the background shrinkage (without a specimen) $L_0$ were performed. Afterwards, the corrected temperature dependence of the shrinkage was calculated using the formula $L(T) = L_{eff}(T) - L_0(T)$ (Fig. 2). This procedure helped to avoid the artifact of "negative shrinkage" (expansion) of the specimens, which was observed in the temperature dependencies of the shrinkage during continuous heating often [42].

The structure investigations were carried out using Jeol® JSM-6490 scanning electron microscope (SEM) with Oxford Instruments® INCA 350 energy dispersion spectroscopy (EDS) X-ray microanalyzer and Jeol® JEM-2100 transmission electron microscope (TEM) with JED-2300 EDS microanalyzer. The Electron Backscatter Diffraction (EBSD) analysis was carried out using Tescan® Vega™ 2 SEM with Nordlys™ 2 analyzer. To identify the alloy microstructure with the use of SEM, electrolytic etching was performed at room temperature in a solution of 75% $H_2SO_4$+15% $HNO_3$+10% HF. The thermal stability of the UFG structure of the Ti alloy was studied by analyzing the structure after 30 min annealing, as well as by step-by-step heating of the specimen *in situ* from 300 °C up to the temperature $T_{ann}$ = 900 °C. The temperature increment was 50 °C, the pause between the heating steps was 30 min.

The microhardness $H_v$ measurements were carried out using Duramin® Struers™ 5 with the load of 2 kg. In order to study the mechanical properties, the relaxation tests were performed to determine the values of the macroelastic limit $\sigma_0$ and the yield strength $\sigma_y$ (see Appendix A in [43]). The rectangular specimens of 3×3 mm in cross-section and 6 mm in height were used in these measurements. The uncertainties of the $\sigma_o$ and $\sigma_y$ measurements were ±30 MPa. The loading rate was 0.13 %/s, the loading time was 0.3 s, the relaxation time was 60 s. The tensile tests of the double blade shaped flat specimens with the working part dimensions of 2×2×3 mm were performed using Tinius Olsen® H25K-S stress machine at the deformation rates ranging from $3.3 \cdot 10^{-3}$ s$^{-1}$.

The electrochemical studies were carried out in the aqueous solution of 10%$HNO_3$ + 0.2%HF using R-8 potentiostat. Based on the analysis of the Tafel slope in the potentiodynamic dependencies [the potential $E$ vs the current density $i$ in the semilogarithmic axes $E - \ln(i)$], the corrosion current density $i_{cor}$ and the corrosion potential $E_{cor}$ were determined. Prior to the electrochemical tests, the specimen surfaces were covered by the corrosion-resistant coating except a 4.5 mm² open area located in the center of the specimen cross-sections in the weld area. Before starting the electrochemical tests, the specimens were kept in the electrochemical cell in the aqueous solution of 0.2%HF+10%HNO$_3$ until a stationary potential value was reached (the holding time was not less than 2 h). Afterwards, the $E(i)$ dependencies were measured at the potential scanning rate of 0.5 mV/s.

### 3. Experimental results

The Ti alloy in the initial state was featured by an irregular grain size distribution. The average grain size varied from 10-20 μm to 50-100 μm (Fig. 3a). The structure of the coarse-grained (CG) alloy included the equiaxed α-phase grains and the elongated α'-phase grains (Fig. 3a, b). The TEM studies of the CG alloy structure revealed a random distribution of the dislocations inside fine α-Ti grains. Some dislocations form the low-angle boundaries. The TEM studies have shown the grain boundaries in the CG alloy to contain the β-phase particles with increased concentration of Fe and V (Fig. 3c-3f). The SEM studies have shown the β-phase particles to be located along the α'-phase platy grain boundaries mainly. The $z$-contrast (the atomic number contrast) analysis has shown the β-phase particles along the grain boundaries to have a different contrast as compared to the one of the α-Ti alloy crystal lattice (Fig. 3b). This observation proved also the β-phase particles to have an increased concentration of the doping elements.

The average grain size after $N = 4$ ECAP cycles was 0.2 to 0.5 μm (Fig. 4). The EDS analysis proved the variations in the local concentration of Al and V along different grain boundaries were insufficient. No precipitation of the β-phase particles along the grain boundaries was found in the UFG alloy.

The *in situ* TEM studies performed during the step-by-step heating demonstrated the recrystallization (the onset of the grain growth) during the annealing of the UFG alloy at 500 ºC and above. Along with the onset of grain growth, the TiC nanoparticles with the average size of 5 to 15 nm were observed in the UFG alloy. Figure 5 shows the precipitation of the TiC particles marked by arrows. After heating up to 700 – 800 ºC, the average size of the TiC particles reached 30 – 50 nm, the average grain size in the UFG alloy was 7 to 9 μm (Fig. 6). No significant differences in the average sizes and in the volume fractions of the TiC particles in the CG and UFG alloys were observed.

The investigations of the mechanical properties proved the formation of the UFG structure in the Ti alloy by ECAP to improve the macroelastic limit $\sigma_0$ from 400 – 420 MPa up to 730 – 750 MPa, the yield stress $\sigma_y$ from 600 – 620 MPa to 1020 – 1050 MPa, and the microhardness limit $H_v$ from 2.0 – 2.1 GPa up to 3.1 – 3.2 GPa, respectively. The investigations of the thermal stability of the mechanical properties demonstrated the softening of the UFG alloy to start after the heating up to 500 – 550 °C (Fig. 7) that corresponded to the recrystallization temperature. A slight increase in the macroelastic limit and yield stress of the Ti alloy at lower temperatures ($T \leq 450 – 500$ °C) results from the precipitation of the TiC nanoparticles apparently.

The stress ($\sigma$) – strain ($\varepsilon$) dependencies for the CG and UFG alloys are presented in Figures 8a and b, respectively. Figure 8c shows the dependencies of the flow stress $\sigma_s$ and of the ultimate elongation to failure $\delta$ on the deformation temperature. The comparative analysis of the tensile mechanical tests at the elevated temperatures has shown the UFG alloy to be featured by a higher plasticity and lower flow stress at high temperatures. At the deformation temperatures increasing from 600 °C up to 800 °C, the flow stress in the CG alloy decreased from 355 – 360 MPa down to 125 – 130 MPa while the plasticity $\delta$ increased from 230% up to 360% (at $T = 700$ °C). The results of testing the UFG alloy specimens in similar conditions demonstrated $\sigma_s$ to decrease from 165 MPa down to 70 MPa while $\delta$ increased from 190% up to 470 – 475% with increasing the deformation temperatures from 600 °C up to 800 °C (Fig. 8c).

Figure 9a shows the dependencies of the shrinkage on the heating time $t$ $L(t)$ for the CG and UFG alloys. Also, Figure 9a shows the temperature curve during heating. Figure 9b presents the dependencies $L(T)$ and $S(T)$ while continuous heating for the CG and UFG alloys. The analysis of the $L(T)$ dependencies revealed a monotonous increase of $L$ during continuous heating, while the $S(T)$ dependencies exhibited two stages with a maximum that is typical for the $S(T)$ dependencies observed during sintering the UFG materials [44, 45]. One can see the shrinkage (the diffusion welding) of the UFG specimens to start at lower temperatures than the one of the CG specimens in similar heating conditions. The degree of shrinkage for the UFG alloy was higher visibly than the one for the CG alloy[1].

The general results of the structural analysis support the observed effect of faster weldability in the UFG alloys. As shown in Figure 10, there was practically no weld visible in the UFG welded specimens. The EBSD analysis helped to locate the weld correctly and to measure the weld width in the welded specimens from the CG alloy. The TEM studies demonstrated the average size and the volume fraction of the pores to be much smaller in the UFG specimens than in the CG ones (Fig. 11). Note that SPS provides very thin welds (Fig. 11b) that ensures much higher operational reliability of the welded structures. The analysis of the histograms in Figures 11c and d showing the pore size distributions (corresponding to Figures 11a and b) revealed the CG metal weld to have bigger pores with an average size of ≈ 1.5 μm, while the UFG one had the pores with the average size of 0.25 – 0.5 μm.

The studies of the mechanical properties demonstrated the hardness of the welds obtained by SPS at the temperatures below the recrystallization threshold to correspond to the hardness of the base metal measured on the base metal out of the weld at the distances of not less than three weld widths from the seam edge (Fig. 12).

Let us analyze the impact of the key factors (the temperature $T$, the pressure $P$, the heating rate $V_h$, and the holding time $t$) on the microstructure parameters of the welded joints obtained by SPS during the high-speed welding of the specimens made from the UFG Ti alloy. The dependencies of $L$ the shrinkage $L$ and of the shrinkage rate $S$ on the above parameters are shown in Figure 13. Figure 13a shows the $L(T)$ and $S(T)$ dependencies (the latter can be also interpreted as the temperature dependence of the welding rate) for the specimens made from the CG and UFG alloys. As one can see in Figure 13a, within the range of $T$ below the $\alpha \leftrightarrow \beta$ phase transition temperature, the maximum shrinkage $L_{max}$ and maximum shrinkage rate $S_{max}$ in the UFG alloy were several times higher than $L_{max}$ and $S_{max}$ in the specimens made from the CG alloy. Within the temperature range above 1000 °C (exceeding the temperature of $\alpha \leftrightarrow \beta$ phase transition), there was no visible difference between the values of $L_{max}$ and $S_{max}$ for the CG and UFG alloys. Thus, one can conclude that at low $T$ the welding intensity in the UFG alloy was much higher than the one in the CG alloy. This was accompanied by a decrease in the typical scale of the diffusion mass transfer $x$, which is proportional to the grain size in the UFG metals: $x = d/2$, and therefore, by a decrease in the standard time of the diffusion mass transfer $\tau_{diff} = \delta D_b/x^3$ where $\delta$ is the grain boundary width and $D_b$ is the grain boundary diffusion coefficient. A small grain size in the UFG Ti alloy reduces $\tau_{diff}$ and, hence, the time required for the diffusion controlled dissolution of the pores located near the non-equilibrium grain boundaries drastically [44].

---

[1] The differences in the initial shrinkage level result from the ≈0.02 mm difference in the specimen heights.

Figure 13b shows the $L_{max}(P)$ and $S_{max}(P)$ dependencies. As one can see in Figure 13b, the increase of $P$ from 50 MPa up to 100 MPa resulted in the increasing of $S_{max}$ from $1.5 \cdot 10^{-3}$ mm/s up to $5.8 \cdot 10^{-3}$ mm/s for the CG alloy and from $1.6 \cdot 10^{-3}$ mm/s up to $23 \cdot 10^{-3}$ mm/s for the UFG alloy. We suppose the results of this study to evidence a significant role of the plastic deformation processes in the diffusion welding of the CG and UFG Ti alloys (for more details, see Discussion section below). This conclusion was supported also by a significant shrinkage of the specimens ($L_{max} = 1 - 8$ mm) exceeding the surface roughness of the specimens after their mechanical polishing ($< 40 - 60$ μm).

The analysis of the $S_{max}(V_h)$ and $L_{max}(V_h)$ dependencies presented in Figure 13c has shown these dependencies for the CG alloy to be descending monotonously; $L_{max}$ decreased from 0.80 mm down to 0.01 mm while $S_{max}$ decreased from $4.0 \cdot 10^{-2}$ mm/s down to $1.5 \cdot 10^{-2}$ mm/s with increasing $V_h$ from 10 °C/min up to 350 °C/min. This effect can be attributed obviously to the decrease of $\tau_{diff}$, within which the diffusion welding of the specimens takes place, and consequently, to the decrease in the time required for the diffusion controlled dissolution of pores in the weld.

The $S_{max}(V)$ and $L_{max}(V)$ dependencies for the UFG alloy were more complex. As one can see in Figure 13c, $L_{max}$ decreased from 2.14 mm down to 1.23 mm with increasing $V_h$ from 10 ºC/min up to 350 ºC/min while $S_{max}$ changed with increasing $V_h$ nonmonotonously. When $V_h$ increases from 10 up to 100 ºC/min, $S_{max}$ increased from $3·10^{-3}$ up to $1.6·10^{-2}$ mm/s. With further increasing of $V_h$ up to 350 ºC/min, $S_{max}$ decreased down to $8·10^{-3}$ mm/s. We suppose, this result can be explained as follows. When $V_h$ grows, the size of the recrystallized grains $d$ decreases leading to the decreasing in typical scale of the diffusion mass transfer $x$ and consequently, to the decrease of the time required to complete the metal sealing and the dissolution of pores in the weld. With further increase of $V_h$, the factor of reduced welding time becomes crucial again and the sealing rate decreases. This conclusion was supported by the results of studying the porosity of the welds in the UFG alloys obtained using diffusion welding at various temperatures and heating rates (Fig. 14, 15).

As it has been shown above (Fig. 12), during the high-rate low-temperature diffusion welding of the UFG alloy, the weld hardness was rather high (3.2 – 3.3 GPa) and approached the weld hardness of the UFG alloy after ECAP (3.5 – 3.6 GPa). Besides, it exceeded the weld hardness achieved by the argon arc welding (2.35 – 2.40 GPa) or by electric fusion welding (2.40 – 2.45 GPa). The microhardness of the weld joints in the CG specimens was 2.4 – 2.6 GPa and was almost independent on the diffusion welding modes.

It is noteworthy that the weld microhardness during welding the UFG alloy specimens at low temperatures (600 – 700 ºC) was less than the one of the base metal away from the weld by 0.4 – 0.6 GPa. Thus, after the diffusion welding of the UFG alloy at 600 ºC ($V_h$ = 100 ºC/min, $P$ = 50 MPa, $t$ = 10 min), the weld microhardness was 2.7 GPa, while the microhardness of the base metal away from the weld was 3.0 – 3.1 GPa. Note also that with increasing $T$ up to 800 ºC, the weld microhradness decreased down to 2.5 GPa, while the metal microhardness away from the weld decreased down to 2.6 – 2.7 GPa. Thus, one can conclude the weld metal obtained after welding the UFG alloy to be featured by a lower hardness than the base metal away from the weld.

Similar results were obtained while studying the impact of $V_h$ and $P$ on $H_v$ in the UFG alloy. The analysis of these results has shown $H_v$ to increase slightly from 2.4 up to 2.6 GPa with increasing $V_h$ from 50 ºC/min to 350 ºC/min ($P$ = 50 MPa, $t$ = 10 min) while $H_v$ of the base metal away from the weld increased from 2.7 – 2.75 up to 2.9 GPa. The increase of $P$ from 50 to 100 MPa ($V_h$ = 100 ºC/min, $t$ = 10 min) resulted in a slight increase of $H_v$ from 2.4 up to 2.5 GPa and to an increase of $H_v$ in the metal away from the weld from 2.7 – 2.8 up to 3.0 – 3.1 GPa.

The values of $H_v$ on the welds in the CG specimens were equal to the ones of the metal away from the weld at $T \leq 700$ ºC and $S \leq 50$ ºC/min was higher than the ones of the metal away from the weld by 0.1 – 0.15 GPa. We attribute the increased $H_v$ of the weld in the CG alloy to the plastic deformation in the surface layers of the CG alloy and, consequently, the strain-induced hardening (work hardening) of this one. Since the defect structure recovery rate in the CG alloy during the high-rate heating is rather low, the hardened layer can survive during the low temperature sintering[2].

The analysis of the weld grain structure parameters has shown the average grain size $d$ in the weld joint to be affected mostly by the welding temperature $T$ and by th eheating rate $V_h$. As one can see in Figure 16, the increase of $T$ from 600 ºC up to 800 ºC ($P$ = 50 MPa, $V_h$ = 100 ºC/min, $t$ =

---

[2] Since the average grain size in the weld in the UFG alloys were almost the same to the one of the metal outside the weld (Fig. 14), the origin of the reduced weld microhardness in the UFG alloys is still unclear. We think the reduced weld microhardness in the UFG alloys to be related most probably to the grain boundary recovery leading to the reduced density of defects in the UFG alloy grain boundaries. The second probable origin of the increased microhardness of the weld-affected zone in the CG alloys may be the creep (as discussed in more details below), which is known to be accompanied by the formation of the dislocation substructures and low-angle boundaries [46, 47]. This hypothesis was confirmed by the results of EBSD analysis of the welds for the CG materials (Fig. 10a), in which the low-angle boundaries with the grain-boundary angle of less than 2º were observed outside the weld predominantly (Fig. 10b).

10 min) resulted in the increase of $d$ in the UFG alloy weld from 3 – 3.8 µm (Fig. 16a) up to 5.9 – 6.9 µm (Fig. 16b). The increase of $V_h$ from 10 ºC/min up to 350 ºC/min ($P$ = 50 MPa, $T$ = 700 ºC, $t$ = 10 min) resulted in the reduction of $d$ in the UFG alloy weld from 6.8 – 7.0 µm down to ≈3.8 µm (Fig. 16c, d). The values of $d$ in the weld joints of the CG specimens corresponded to the ones in the CG alloy within the measurement uncertainty.

We have compared the $d(T)$ dependencies ($t$ = 10 min) to the ones for the CG and UFG alloys during the 30-min annealing [14] and found that during the diffusion welding, the grain growth started at lower temperatures while the intensity of the grain resizing during the diffusion welding at the elevated temperatures appeared to be much lower (Fig. 17). At high $T$, this allows obtaining the homogeneous fine-grained structure with smaller grain size unlike ordinary annealing (Fig. 17). Note also that the same effect was observed in both CG and UFG alloys.

We suppose the obtained result to prove the recrystallization (the grain growth) mechanisms during the diffusion welding to differ from the recrystallization mechanisms during annealing of the UFG alloy. The preserved equiaxial grain structure and weak grain growth may indicate that the recrystallization process during the diffusion welding of the CG and UFG alloys, in the first approximation, can be presented as a competition of two processes:

(i) the formation of the submicron fragments ("recrystallization nuclei") during the high-temperature plastic deformation, and

(ii) fast migration of the fragment boundaries.

Let us analyze the corrosion resistance in the weld joints obtained.

When comparing the potentiodynamic dependencies (Fig. 18) for the CG and UFG alloys after the diffusion welding, we noted the values of $i_{cor}$ for the welds from the UFG alloy to be lower than the one in the CG alloy welds obtained at similar values of $T$ and $V_h$ (Table 2). Note that at higher $T$, the differences in the corrosion resistances between the CG and UFG alloy welds increased. After welding at 600, 700, and 800 ºC, the values of the ratio of the $i_{cor}$ values for the UFG alloy welds and the CG ones were 1.17, 1.84, and 1.97, respectively. Note that the $i_{cor}(T)$ dependencies were manifested `differently in the CG and the UFG alloys (Table 2). The nonmonotonous $i_{cor}(T)$ dependence with the maximum at $T \approx 700$ °C has been observed during welding the CG alloy specimens in the α-phase temperature range (Table 2). For the UFG alloy, $i_{cor}$ decreased monotonously with increasing $T$ in the α-phase temperature range. The corrosion potential for the UFG alloy was more negative (by 10 – 15 mV) as compared to the one for the CG one. Diffusion welding in the β-phase area resulted in the formation of the martensite structure (Fig. 19) and reduced the corrosion resistance of the welds (Table 2).

The analysis of the potentiodynamic dependencies has shown the higher heating rates to result in a bigger corrosion resistance of the welds. The values of $i_{cor}$ in the UFG alloy welds obtained at $V_h$ = 50 °C/min was 1.32 mA/cm², while at $V$ = 350 °C/min $i_{cor} \approx 0.18$ mA/cm². The values of $i_{cor}$ for the CG alloy welds obtained at $V_h$ = 50 and 350 °C/min were 1.51 and 0.99 mA/cm², respectively. Note that the $i_{cor}(V_h)$ dependencis (welding at $T$ = 700 °C) were nonmonotonous with the maxima at $V_h \approx 100$ °C/min for the CG alloy and $V_h \approx 50$ °C/min for the UFG ones.

The analysis of the dependence of the corrosion potential on the holding time $t$ (Fig. 18b, see also Tables 2 and 3) has shown the steady potential for the UFG alloy welds to be somewhat less than the one for the CG ones that also confirms the increased corrosion resistance of the UFG alloy welds.

Thus, one can conclude the corrosion resistance of the UFG alloy welds to be higher than the one of the CG alloy welds.

Figures 20a and b show the images of the surfaces of the CG and UFG alloy specimens, respectively after the corrosion electrochemical investigations. As one can see in Figure 20, rather active corrosion decay occurred in the weld during the electrochemical investigation. Also, it is evident from Figure 20 that etching (corrosion decay) was observed during the testing at the grain boundaries of the Ti alloys. This suggests any changes observed in the corrosion properties to be primarily due to the processes occurring at the grain boundaries during the high-rate diffusion welding.

We ascribe the increased corrosion resistance of the UFG alloy to the changes in the structural-phase composition of the grain boundaries in the Ti alloy during ECAP (Fig.4). As it has been mentioned above, ECAP performed at elevated temperatures (450 – 475 ºC) resulted in reduced concentrations of the detrimental corrosive element (V) along the grain boundaries and in reduced differences in the concentrations of V ($\Delta C_V$) and Al ($\Delta C_{Al}$) in the crystal lattice $C_v$ and at the grain boundaries $C_b$: $\Delta C_V = C_{v(V)} - C_{b(V)}$, $\Delta C_{Al} = C_{v(Al)} - C_{b(Al)}$. The differences in the concentrations of Al and V between the crystal lattice and at the grain boundaries in the CG Ti - 5Al-2V alloy may reach $\Delta C_{Al}$ = 5.6 at.% and $\Delta C_V$ = 12.9 at.%, respectively. The studies of the grain boundary composition in the UFG alloy after ECAP revealed $\Delta C_{Al}$ and $\Delta C_V$ to be 0.35 – 0.40 at.% that exceeds slightly the range of the experimental uncertainty of determining the Al and V concentrations by EDS (± 0.2 – 0.3 at.% [14]).

While heating the UFG alloy to the temperatures exceeding the recrystallization threshold, fast migrating grain boundaries capture the atoms of the doping elements (Al and V) distributed in the crystal lattice uniformly that leads to the increasing of $\Delta C_{Al}$ and $\Delta C_V$ again. In this case, the concentration of doping elements at the grain boundaries will obviously be proportionate to the typical distance over which the grain boundaries move and, consequently, to the size of the recrystallized grains. Thus, a decrease in the migration rate of the grain boundaries due to the increased $S$ or decreased $T$ was expected to boost the corrosion resistance of the Ti alloys. This effect was observed in the experiment indeed (see Table 3).

## 5. Discussion

Let us analyze the weld sealing parameters during the high-speed diffusion welding of the CG and UFG Ti alloys. As mentioned above, $L(T)$ and $S(T)$ dependencies for the CG alloys are featured by two stages usually while the *L*(*T*) and *S*(*T*) dependencies for the UFG alloys have more complex multistage character (Fig. 9b).

First, it is noteworthy that the temperature range $T = 600 - 800$ °C (that makes $(0.44 - 0.56)T_m$, where $T_m = 1933$ K is the α-Ti melting point) and the stress range (50 – 100 MPa that makes $(1.1 - 2.3) \cdot 10^{-3}$ *G*, where $G = 43.6$ GPa is the shear modulus of α-Ti) of the diffusion welding in the M.F. Ashby deformation mechanism maps [48] correspond to the exponential creep range (Stage III in the deformation mechanism maps). According to [48], the dependence of the strain rate ($\dot{\varepsilon}$) on applied pressure ($\sigma_s$) and the test temperature for Stage III can be described by the following equation: $\dot{\varepsilon}_1 = A_s D_{eff}(Gb/kT)(\sigma_s/G)^n$, where $A_s$ is a numerical coefficient, $D_{eff}$ is the effective diffusion coefficient, *b* is the Burgers vector, and *k* is the Boltzmann constant. Based on mass empirical generalization, M.F. Ashby had shown [48] the exponential creep activation energy may correspond to the grain boundary diffusion activation energy $Q_b = 97$ kJ/mol, $D_{eff} = \delta D_b$ or to the bulk diffusion activation energy $Q_v = 242$ kJ/mol, $D_{eff} = D_v$ subject to the grain sizes.

To identify the creep mechanism during the diffusion welding of the Ti alloy specimens, the dependencies of the shrinkage *L* on the isothermic time *t* at 700 °C under different stresses (50, 70, and 100 MPa) were analyzed[3]. These *L*(*t*) dependencies for the CG and UFG specimens are presented in Figures 21a and b. As one can see from these data, the stage of steady flow was observed in the *L*(*t*) dependencies for all alloys. Based on the assumption that at this stage the shrinkage is proportional to the deformation of the specimen, the usual procedure of plotting the $\lg(\sigma_s) - \lg(\dot{L})$ dependence can be used to estimate the index *n*, while the creep activation energy $Q_{cr}$ can be determined by the slope of the $\ln(\dot{L}) - T_m/T$ dependence.

---

[3] The weld specimen was heated at the same rate (100 °C/min) up to 700 °C, then was held at this temperature for 10 min under different values of the uniaxial pressure $\sigma_s = 50, 70$, and 100 MPa (Fig. 21).

This procedure revealed the index $n$ determined from the slope of the $\lg(\sigma_s) - \lg(\dot{L})$ dependence for the CG alloy at 700 °C (approximately $0.51 T_m$) to be $5.6 \pm 0.8$ (Fig.22a). This value exceeds the tabular value $n = 4.3$ published for α-Ti [46] slightly that can be related to the impact of the β-phase particles on the creep rate in the Ti alloy leading to significant increase of $n$ [49, 50]. The analysis of the $L(t)$ dependencies for different temperatures (600 °C, 700 °C, and 800 °C) under a constant pressure ($P = 50$ MPa) revealed the value of $Q_{cr}$ for the CG Ti alloy to be $\approx 14.9\ kT_m$ ($\approx 240$ kJ/mol, Fig. 22b) that appears to be close to the activation energy of the exponential creep, the rate of which is limited by the α-Ti lattice diffusion intensity [48].

Similar analysis of the $L(t)$ dependencies under different stresses (50, 70, and 100 MPa) at different temperatures (600, 700, and 800 °C) has shown the values of $Q_{cr}$ in the UFG alloy during the isothermic holding to be $\approx 6.2\ kT_m$ ($\approx 100$ kJ/mol, Fig. 22b). This value of $Q_{cr}$ is close to the grain boundary diffusion activation energy $Q_b$ and to the lattice dislocation nuclei activation energy $Q_c$ in α-Ti ($Q_b = Q_c = 97$ kJ/mol [48]). The index $n$ determined from the slope of the $\lg(\sigma_s) - \lg(\dot{L})$ dependence for the UFG alloy was $2.1 - 2.7$ (Fig. 22a). The values of $Q_{cr}$ are in good agreement with the literature data on the creep flow in the UFG Ti and Ti alloys [51, 52]. On the other hand, the values of $n$ and $Q_{cr}$ obtained for the UFG Ti alloy are in good agreement with the values for the exponential creep, the rate of which is limited by gliding and climb (creep) of dislocations [48].

When studying the thermal stability of the UFG alloy, the nanoparticles of the second phase (presumably TiC) were found to precipitate. These ones were expected to hinder the movement of the lattice dislocations and to increase the index $n$ in the exponential creep equation. According to the results of the above analysis, the values of the index $n$ in the UFG alloy appear to be lower than the theoretical value $n_{th} = 3$ for the exponential creep mechanism [48], below $n = 4.3$ typical for the CG Ti [48], and much below the values of $n$ observed in the UFG Ti [53, 54].

Note that some authors [53, 54] attributed the low values of the $n$ observed in the UFG metals and low values of $Q_{cr}$ comparable to the grain boundary diffusion activation energy $Q_b$ rto the superposition of several mechanisms of plastic deformation with low and high values of $n$ during the creep tests. In the case of the plastic deformation of the UFG alloys at relatively low temperatures, the Coble creep ($\dot{\varepsilon}_2 = B_s(\delta D_b/d^3)(\sigma_s b^3/kT)$ [55]) or the grain boundary sliding ($\dot{\varepsilon}_3 = A_b(\delta D_b/kT)(b/d)^2(\sigma/G)^2$ [56, 57], where $B_s$ and $A_b$ are the numerical coefficients, $\delta = 2b$ may occur.

The analysis of the results of the mechanical tensile tests performed at elevated temperatures allows determining the strain rate sensitivity coefficient $m$ from the slope of the dependencies of the flow stress $\sigma_s$ on the deformation rate $\dot{\varepsilon}$ in the logarithmic axes $m = \partial \ln(\sigma^*)/\partial \ln(\dot{\varepsilon})$. The analysis of the $\sigma_s(\dot{\varepsilon})$ dependencies presented in Figure 23 has shown $m$ in the CG titanium alloy at $T = 700$ °C to be $0.24 - 0.28$ while in the UFG alloy $m = 0.50 - 0.56$. High values of $m$ in the UFG alloy evidence a significant contribution of the grain boundary sliding mechanism into the total elongation of the UFG alloy specimens during the deformation at elevated temperatures.

Thus, one can assume two simultaneous processes to take place during the isothermic weld sealing in the UFG Ti alloy:

(i) the exponential creep associated with the gliding and climb of dislocations, the activation energy of which corresponds to the diffusion activation energy in the lattice dislocation nuclei, and

(ii) the grain boundary sliding or Coble creep, the activation energy of which corresponds to the grain boundary diffusion activation energy.

Let us analyze the weld sealing patterns in the CG and UFG alloys during continuous heating.

The analysis of the dominant diffusion mechanisms observed at early sintering stages during continuous heating can be carried out using the Young-Cutler model [58] for the non-isothermic sintering modes. According to this model, the effective sintering activation energy in the continuous heating conditions $mQ_s$ where $m$ is a numerical coefficient ($m = 1.2$ in the case of the diffusion in the bulk crystal lattice and $m = 1/3$ in the case of the grain boundary diffusion) [58-60] can be determined from the slope of the $\ln(Td\varepsilon/dT) - T_m/T$ dependence [58], where $\varepsilon = L/L_0$.

Figure 24 shows the $L(T)$ dependencies for the CG and UFG alloys in the $\ln(Td\varepsilon/dT) - T_m/T$ axes. The analysis of the obtained results has shown that the above dependencies can be approximated by a straight line with a reasonable degree of accuracy. The values of $mQ_s$ for the CG weld sealing didn't depend on $S$ within the experimental uncertainty and equals to $(8.2 - 9.3)kT_m$. If $m = 1/2$, the CG alloy sealing activation energy during the continuous heating makes $(16.4 - 18.6)kT_m$ (265 – 298 kJ/mol), which is rather close to the exponential creep activation energy in α-Ti limited by gliding and climb of dislocations ($Q_{cr}$ = 242 kJ/mol [48]).

The values of $mQ_s$ in the UFG Ti alloy appeared to depend on $S$: $mQ_s$ decreased from $(4.1 \pm 0.3)kT_m$ down to $(2.2 \pm 0.5)kT_m$ with increasing $S$ from 10 °C/min to 100 °C/min. Further increase of $S$ up to 350 °C/min resulted in a slight increase of $mQ_s$ up to $(3.2 \pm 0.4)kT_m$. If $m = 1/3$, the values of $Q_s$ $(6.6 - 12.3)kT_m$ = 106 – 197 kJ/mol are close to the ones for the grain boundary diffusion in pure α-Ti ($Q_b$ = 97 kJ/mol) and to the recrystallization activation energy in the UFG alloy ≈ $6.7kT_m$ ≈108 kJ/mol [61]. Higher activation energy values (106 – 197 kJ/mol) in the UFG alloy Ti-5Al-2V were ascribed to the impact of the doping elements on the diffusion permeability of the grain boundaries in the UFG Ti alloy. As has been shown above (Fig. 4c), the grain boundaries in the UFG Ti alloy contain an enhanced concentration of Al and V (see also [14, 61]).

Note that the nonmonotonous character of the dependence of the weld sealing activation energy in the UFG alloy corresponds to the nonmonotonous $S(V_h)$ dependence. This fact can be considered as an indirect proof of the analytical method chosen to be correct.

Thus, one can conclude that the high rates of the diffusion welding of the UFG Ti alloy result from the small grain sizes that reduces typical scale of the diffusion mass transfer significantly as well as from the change in the diffusion mechanism: the diffusion in the crystal lattice is replaced by the diffusion along the grain boundaries, the latter being much more intensive than the former. This allows using the SPS technology for making the high-density fine-grained structure in the weld featured by high density and corrosion resistance.

## 6. Conclusions

1. SPS is shown to allow efficient high-speed diffusion welding of α-Ti alloys and, at the same time, to preserve the UFG structure, high hardness, and corrosion resistance in the weld. The pores in the UFG alloy welds were found to diffuse faster due to highly intensive diffusion processes taking place along the non-equilibrium grain boundaries and the small-scale diffusion mass transfer, typical scale of which is proportional to the grain size.

2. The dependence of the weld sealing rate (welding rate) on the heating rate was shown to be different for the CG and UFG alloys. An increase in the heating rate in the CG alloys leads to a monotonous decrease of the sealing rate caused by the reduced welding time at high heating rates. The weld sealing rate in the UFG alloys depended on the heating rate nonmonotonously (with a maximum) that was attributed to the decrease of the recrystallized grain size with increasing heating rate leading to the decrease in the typical scale of diffusion mass transfer and, consequently, to the decrease of the standard time required to complete the metal sealing and to the dissolution of the pores in the weld. The factor of the reducing the welding time becomes crucial again with increasing heating rate, and the sealing rate decreases.

3. At the stage of the continuous heating, the kinetics of the weld sealing in the CG and UFG Ti-5Al-2V alloys can be described by the Young-Cutler model. The weld sealing activation energy in the CG Ti alloy during the continuous heating was close to the exponential creep activation energy in α-Ti limited by gliding and climb of dislocations. The weld sealing activation energy during the diffusion welding of the UFG alloy was close to the grain boundary diffusion activation energy.

4. At the stage of isothermic holding, the kinetics of the weld sealing in the Ti alloys is limited by the exponential creep processes, the rate of which in the CG Ti alloy is determined by the intensity of diffusion in α-Ti crystal lattice. In the UFG alloy, the exponential creep processes associated with gliding and climb of dislocations, the activation energy of which corresponds to the diffusion activation energy in the lattice dislocation nuclei may take place simultaneously with the grain boundary sliding and Coble creep, the activation energy of which corresponds to the grain boundary diffusion activation energy.

**Acknowledgements**

The authors thank A.V. Piskunov and N.V. Sakharov (Lobachevsky Univ.) for developing the methods of EBSD analysis of the titanium alloy welds. The authors thank E.A. Lantsev (Lobachevsky Univ.) for conducting the tests in order to measure the temperature-shrinkage dependencies $L_0(T)$ without a specimen using Dr. Sinter® SPS-625 setup in different heating modes.

The authors thank Afrikantov OKB Mechanical Engineering JSC for performing the argon-arc and electron-beam welding of the UFG alloy specimens.

**Data Availability**

The raw/processed data required to reproduce these findings cannot be shared at this time as the data also form a part of an ongoing study.

**Compliance with ethical standards**

**Conflict of interest**. The authors declare that they have no conflict of interest.

**Table 1. Composition of the titanium alloy (wt.%)**

|  | Ti | Al | V | Zr | Fe | Si | $O_2$ | $N_2$ | C | $H_2$ |
|---|---|---|---|---|---|---|---|---|---|---|
| Tested alloy | Balance | 4.73 | 1.88 | 0.019 | 0.11 | 0.03 | 0.042 | 0.01 | 0.0024 | 0.004 |
| Russian standard GOST 19807-91 | Balance | 3.5-5.0 | 1.2-2.5 | ≤0.30 | ≤0.25 | ≤0.12 | ≤0.15 | ≤0.04 | ≤0.1 | ≤0.006 |

**Table 2.** Results of electrochemical studies of the welds from the CG and UFG alloys ($V_h$ = 50 °C/min, $P$ = 50 MPa, $t$ = 10 min). Analysis of the impact of the diffusion welding temperature on the corrosion resistance

| State | SPS welding temperature $T$ (°C) | Corrosion current density $i_{cor}$ (mA/cm$^2$) | Corrosion potential $E_{cor}$, mV |
|---|---|---|---|
| CGalloy | 600 | 1.21 | -471 |
| | 700 | 1.69 | -500 |
| | 800 | 1.30 | -456 |
| | 1030$^{(*)}$ | 1.74 | -467 |
| UFG alloy | 600 | 1.03 | -486 |
| | 700 | 0.92 | -507 |
| | 800 | 0.66 | -468 |
| | 1140$^{(*)}$ | 0.82 | -499 |

$^{(*)}$ diffusion welding in the β-phase temperature range

**Table 3. Results of electrochemical studies of the welds from the CG and UFG alloys ($T = 700$ °C, $P = 50$ MPa, $t = 10$ min). Analysis of the impact of the heating rate during the diffusion welding on the corrosion resistance**

| State | Heating rate $V_h$ (°C/min) | Corrosion current density $i_{cor}$ (mA/cm²) | Corrosion potential $E_{cor}$, mV |
|---|---|---|---|
| CG alloy | 10 | 1.12 | -473 |
|  | 50 | 1.51 | -476 |
|  | 100 | 1.69 | -500 |
|  | 350 | 0.99 | -509 |
| UFG alloy | 10 | 0.66 | -500 |
|  | 50 | 1.32 | -479 |
|  | 100 | 0.92 | -507 |
|  | 350 | 0.16$^{(*)}$ | -180$^{(*)}$ |

$^{(*)}$ specimen while holding in the electrolytic solution remained in the passive state (the surface corrosion was not observed)

**List of figures**

**Figure 1.** Schematic representation of the Spark Plasma Sintering apparatus for high-rate diffusion welding.

**Figure 2**. Dependence of the shrinkage on the heating temperature in the CG specimen of the titanium alloy: (a) thermal expansion contribution accounting procedure (curve (1) – measured (effective) shrinkage; line (2) – apparatus contributed thermal expansion; line (3) – true shrinkage; (b) impact of the initial roughness on the shrinkage of specimens during continuous heating (line (1) – roughness 3–5 µm, line (2) – roughness 20–28 µm, line (3) – roughness 40–60 µm).

**Figure 3**. Structure of the coarse-grained titanium alloy: (a, b) – alloy microstructure; (c–f) – precipitation of the β-phase particles (TEM) (β-phase particles are shown with arrows); (e–f) EDS results for the grain boundaries composition.

**Figure 4**. Microstructure of the UFG titanium alloy: (a) bright-field TEM image; (b) dark-field TEM image; (c) electron diffraction pattern; (d) EDS results for the grain boundaries composition.

**Figure 5**. Precipitation of the titanium carbide particles while studying the UFG titanium alloy by TEM while heating up *in-situ*. Precipitated particles are marked by arrows. Heating the UGF alloy up to 600 ºC and holding 30 min. (a); heating up to 700 ºC and holding for 30 min. (b).

**Figure 6**. SEM micrographs of the UFG titanium alloy after annealing at 600 ºC, 30 min (a) and 700 ºC, 30 min (b).

**Figure 7**. Dependence of the microhradness (circles), macroelastic limit (squares), and yield strength (rhombus) on the annealing temperature (30 min) for the CG (light markers) and UFG (dark markers) titanium alloy specimens.

**Figure 8**. Results of mechanical testing at elevated temperatures: (a–b) stress ($\sigma$) – strain ($\varepsilon$) curves for the CG (a) and UFG titanium alloy (b); (c) dependencies of yield stress ($\sigma_s$) and of ultimate elongation to failure ($\delta_f$) on the deformation temperature for the CG alloy (1) and for UFG alloy (2). Deformation rate $3.3 \cdot 10^{-2}$ s$^{-1}$.

**Figure 9**. Shrinkage curves (welding diagrams) for the CG and UFG alloy specimens ($V$ = 100 °C/min, $P$ = 100 MPa): (a) dependence of temperature and shrinkage on the heating time for CG (1) and for UFG (2) alloy; (b) dependences of shrinkage and shrinkage rate on the welding temperature for CG (1) and UFG (2) alloy.

**Figure 10**. Microstructure of the metal in the weld obtained during the diffusion welding of CG (a) and UFG (b) alloys. EBSD analysis. The low-angle boundaries are marked by white.

**Figure 11**. A weld from the CG (a) and UFG (b) alloys made at the same welding temperature ($T$ = 850 °C), heating rate ($V_h$ = 50 °C/min), and pressure ($P$ = 50 MPa); the histograms of pore size distributions in the CG (c) and UFG (d) alloys. White arrows mark the pores.

**Figure 12**. Distribution of microhardness in the cross sections of the welds from UFG alloy made by different welding technologies: (1) diffusion welding using SPS ($T$ = 600 °C, $V$ = 100 °C/min, $P$ = 100 MPa, $t$ = 0 min); (2) argon-arc welding, (3) electron-beam welding ($h$ is the distance from the center of the weld seam).

**Figure 13**. Dependencies of the maximum shrinkage $L_{max}$ and maximum shrinkage rate $S_{max}$ during the high-rate diffusion welding of the CG (1) and UFG (2) titanium alloys: (a) impact of welding temperature ($V = 100$ °C/min, $P = 50$ MPa); (b) impact of the applied pressure ($T = 700$ °C, $V = 100$ °C); (c) impact of the heating rate ($T = 700$ °C, $P = 50$ MPa).

**Figure 14**. Microstructure of the UFG alloy welds: $V = 100$ °C/min, $P = 70$ MPa, $t = 10$ min, $T = 600$ °C; (b) $V = 10$ °C/min, $P = 50$ MPa, $t = 10$ min, $T = 700$ °C. The arrows mark the weld.

**Figure 15**. Histograms of the pore size distributions in the UFG alloy weld in the frequency ($N/N_\Sigma$) – pore size $R$ axes: (a) $T = 600$ °C, $V = 100$ °C/min, $P = 50$ MPa, $t = 10$ min; (b) $T = 800$ °C, $V = 100$ °C/min, $P = 50$ MPa, $t = 10$ min; (c) $T = 700$ °C, $V = 10$ °C/min, $P = 50$ MPa, $t = 10$ min; (d) $T = 700$ °C, $V = 350$ °C/min, $P = 50$ MPa, $t = 10$ min.

**Figure 16**. Microstructure of the UFG alloy welds: (a) $V = 100$ °C/min, $P = 50$ MPa, $t = 10$ min, $T = 600$ °C; (b) $V = 100$ °C/min, $P = 50$ MPa, $t = 10$ min, $T = 800$ °C; (c) $V = 10$ °C/min, $P = 50$ MPa, $t = 10$ min, $T = 700$ °C; (d) $V = 350$ °C/min, $P = 50$ MPa, $t = 10$ min, $T = 700$ °C.

**Figure 17**. Dependence of the average grain size on the annealing temperature (30-min) (black markers) (see [14]) and dependence of the average grain size on the temperature of the diffusion welding (white markers) for the CG (square markers) and UFG alloy (round markers).

**Figure 18**. Potential – current density (a) and potential – holding time (b) dependencies for the CG (1–4) and UFG (5–8) titanium alloy welds formed using different regimes of the diffusion welding by SPS: (1), (3) 600 °C, 100 °C/min, 10 min, 50 MPa; (2), (4) 800 °C, 100 °C/min, 10 min, 50 MPa; (5, 6) 700 °C, 10 °C/min, 10 min, 50 MPa; (7, 8) 700 °C, 350 °C/min, 10 min, 50 MPa.

**Figure 19**. Microstructure of the CG (a) and UFG (b) titanium alloy after diffusion welding at the temperature exceeding the temperature of (α↔β) phase transition

**Figure 20**. Surface of CG (a) and UFG (b) alloy welds after corrosion electrochemical investigation. The CG and UFG alloy specimens were obtained in similar welding regimes ($V = 10$ °C/min, $P = 50$ MPa, $t = 10$ min, $T = 700$ °C)

**Figure 21**. Dependences of shrinkage on the isothermic time for the CG (a) and UFG (b) titanium alloy welds under different applied pressure (test temperature 700 °C)

**Figure 22**. Dependences of shrinkage rate on the stress in the $\lg(\dot{L}) - \lg(\sigma_s)$ axes (test temperature 700 °C) (a) and dependences of shrinkage rate on the reciprocal homologic temperature in $\lg(\dot{L}) - T_m/T$ axes (test pressure 50 MPa) (b)

**Figure 23**. The analysis of the dependences of flow stress on the deformation rate for the CG (1) and UFG (2) alloys shown in Fig.8. Deformation temperature 700 °C. dependencies shown in Fig.8

**Figure 24**. Analysis on the basis of the Young-Cutler model of dependencies of shrinkage on heating temperature during continuous heating ($V = 100$ °C/min, $P = 50$ MPa) for the weld seals in the CG and UFG alloys in the $\ln(Td\varepsilon/dT) - T_m/T$ axes: (1) CG alloy ($V = 100$ °C/min, $P = 50$ MPa); (2) UFG alloy ($V = 10$ °C/min, $P = 50$ MPa); (3) UFG alloy (V = 100 °C/min, P = 50 MPa)

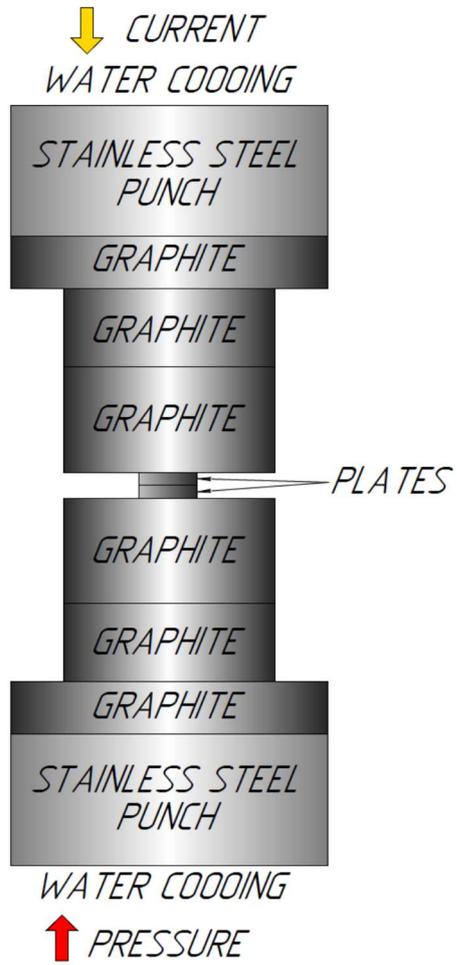

**Figure 1**

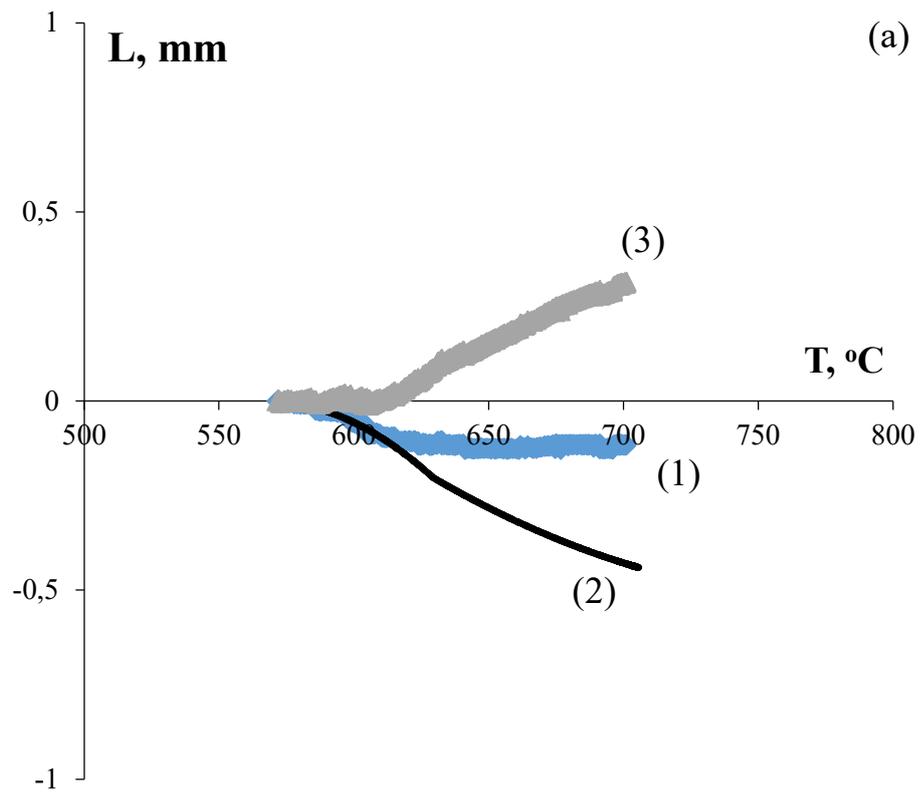

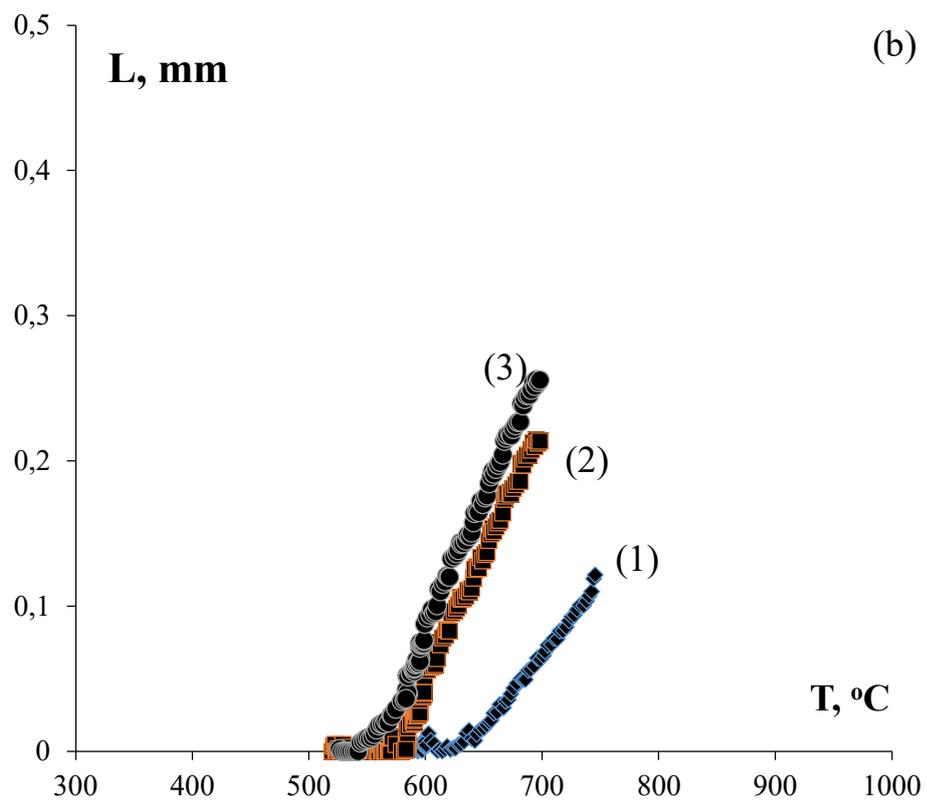

**Figure 2**

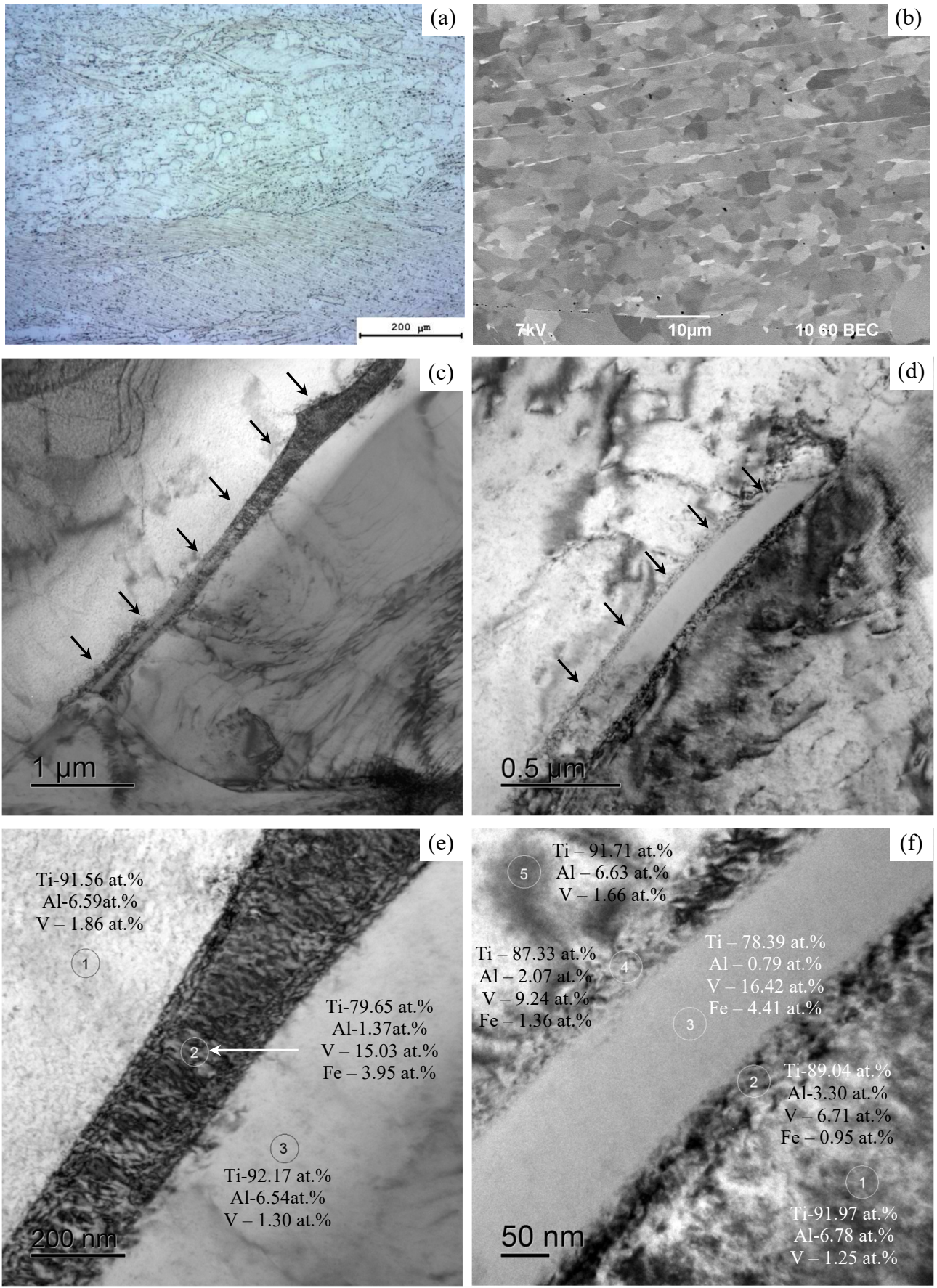

**Figure 3**

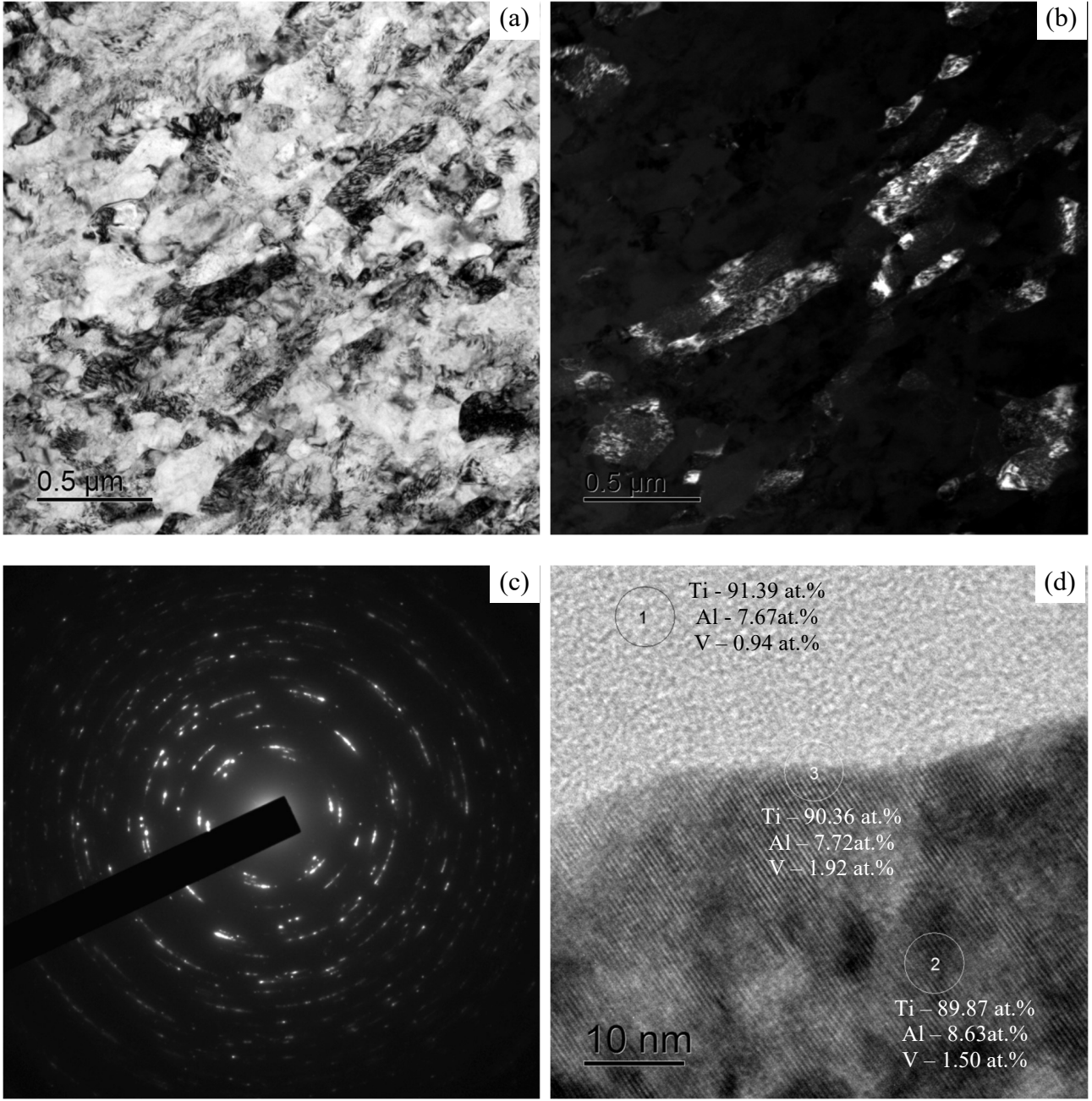

**Figure 4**

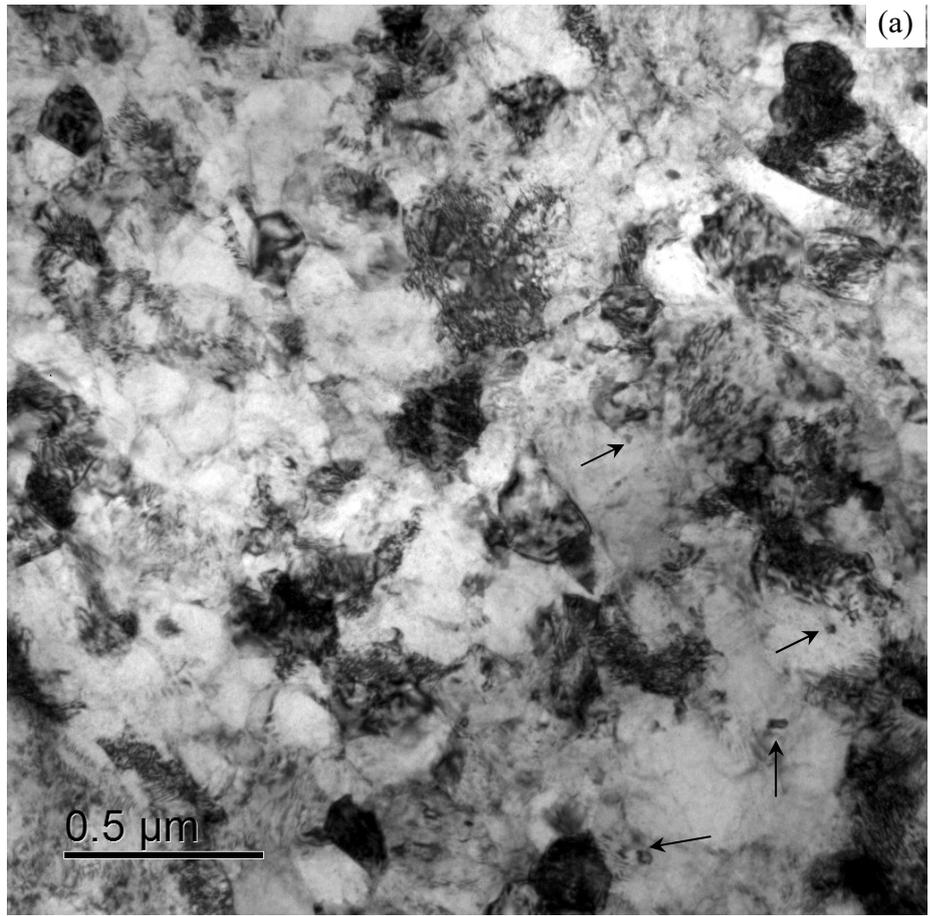
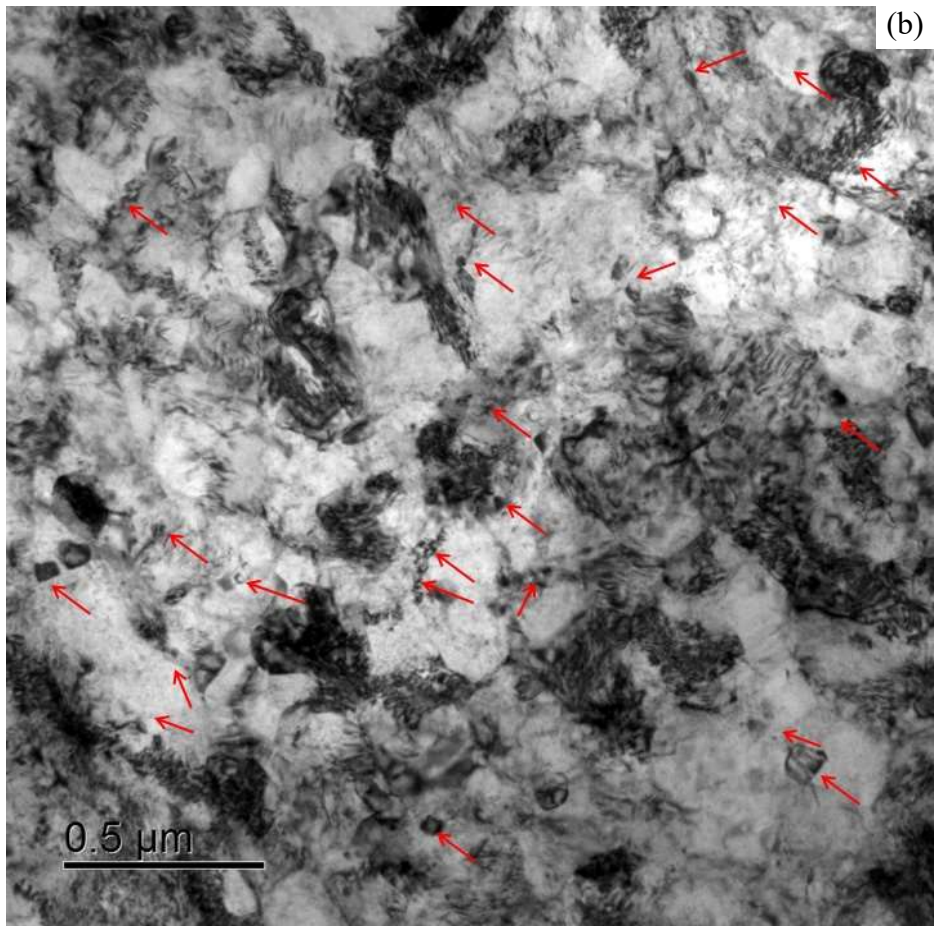

**Figure 5**

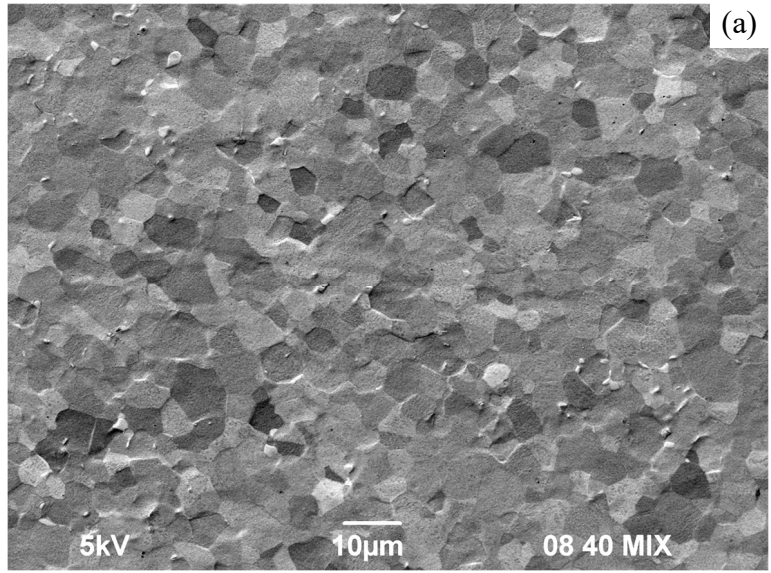
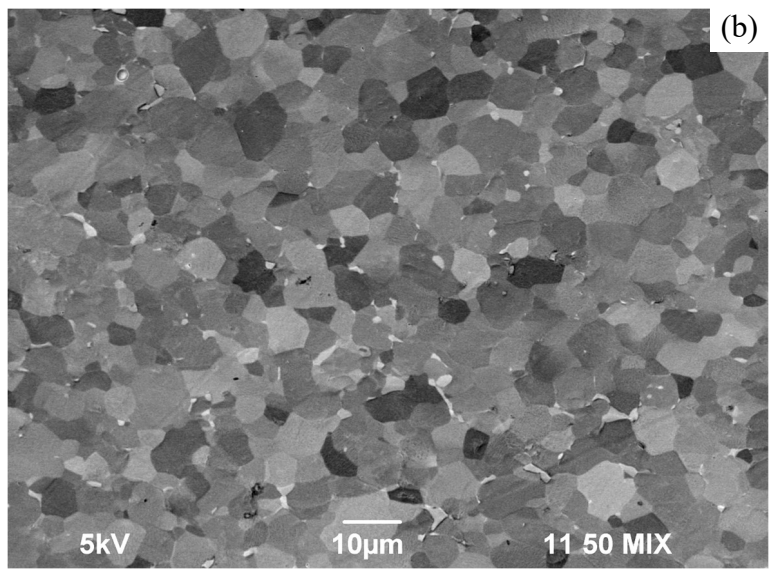

**Figure 6**

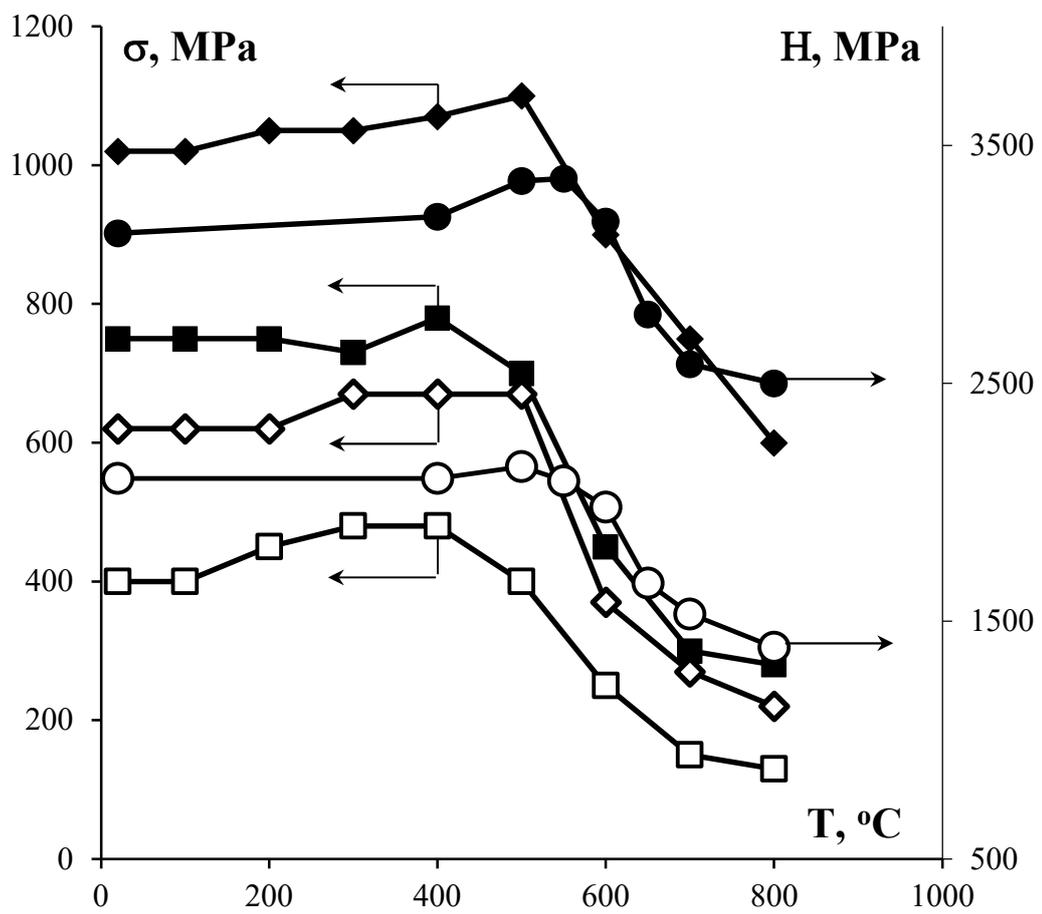

**Figure 7**

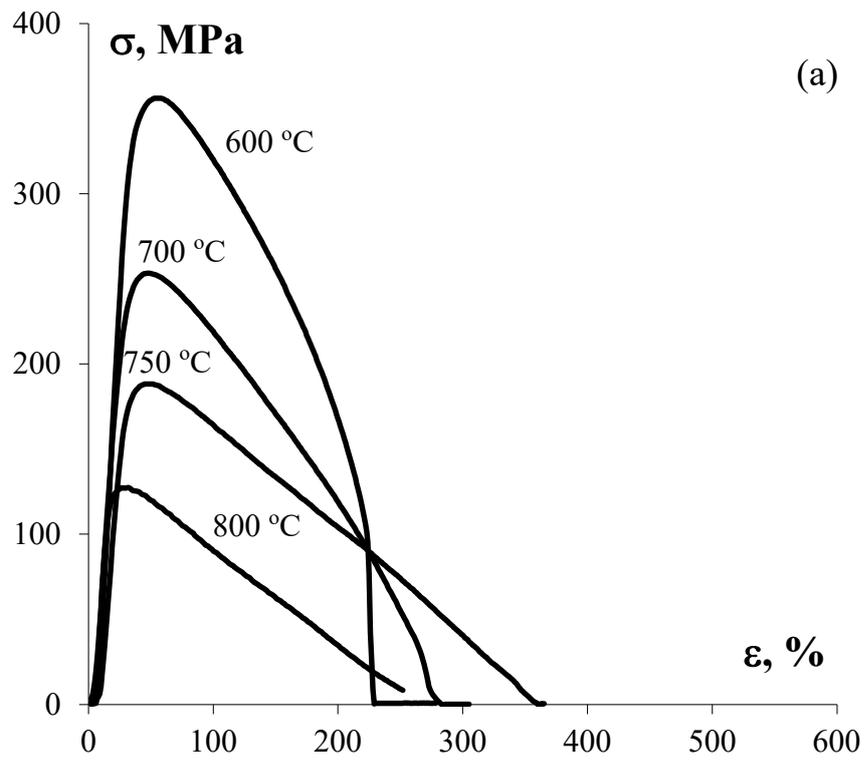

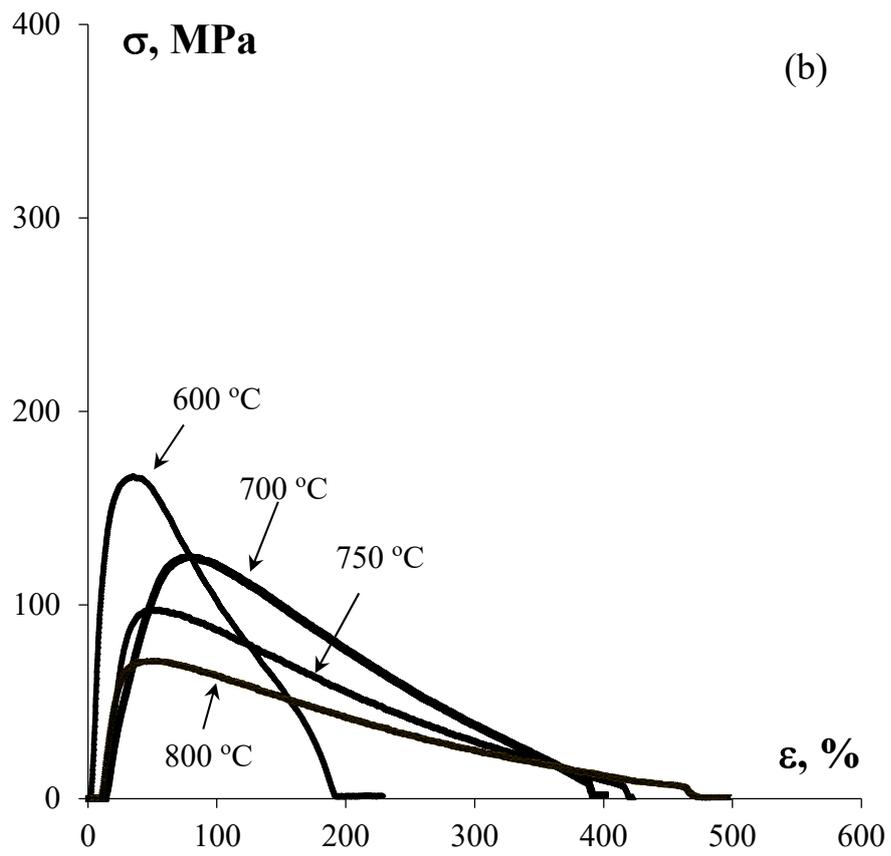

**Figure 8a, b**

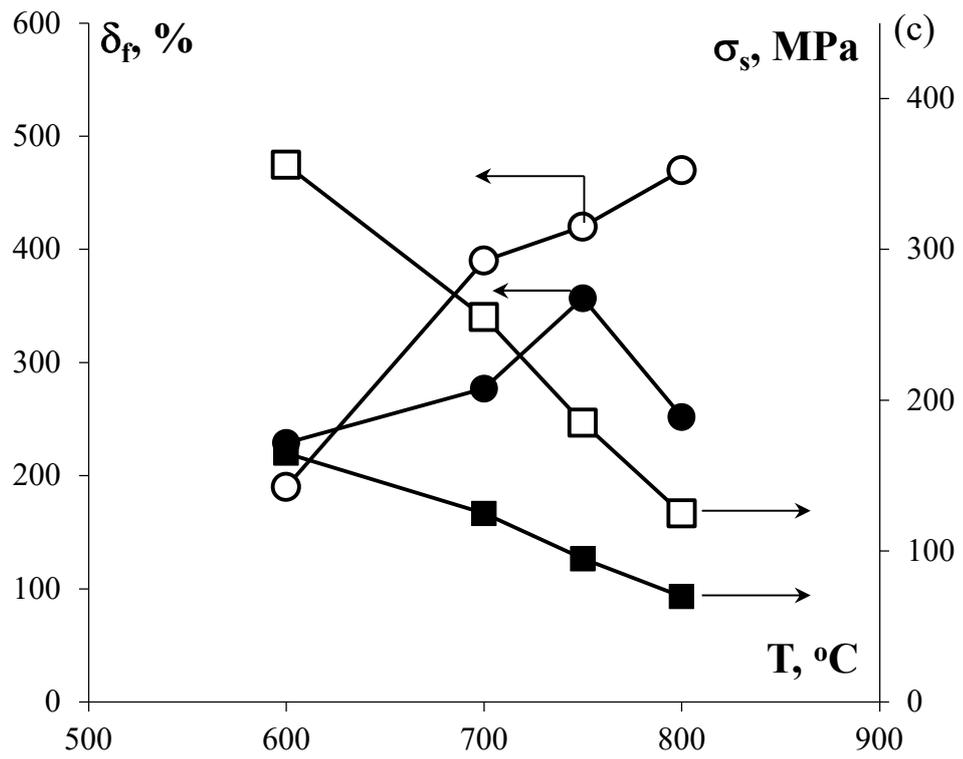

**Figure 8c**

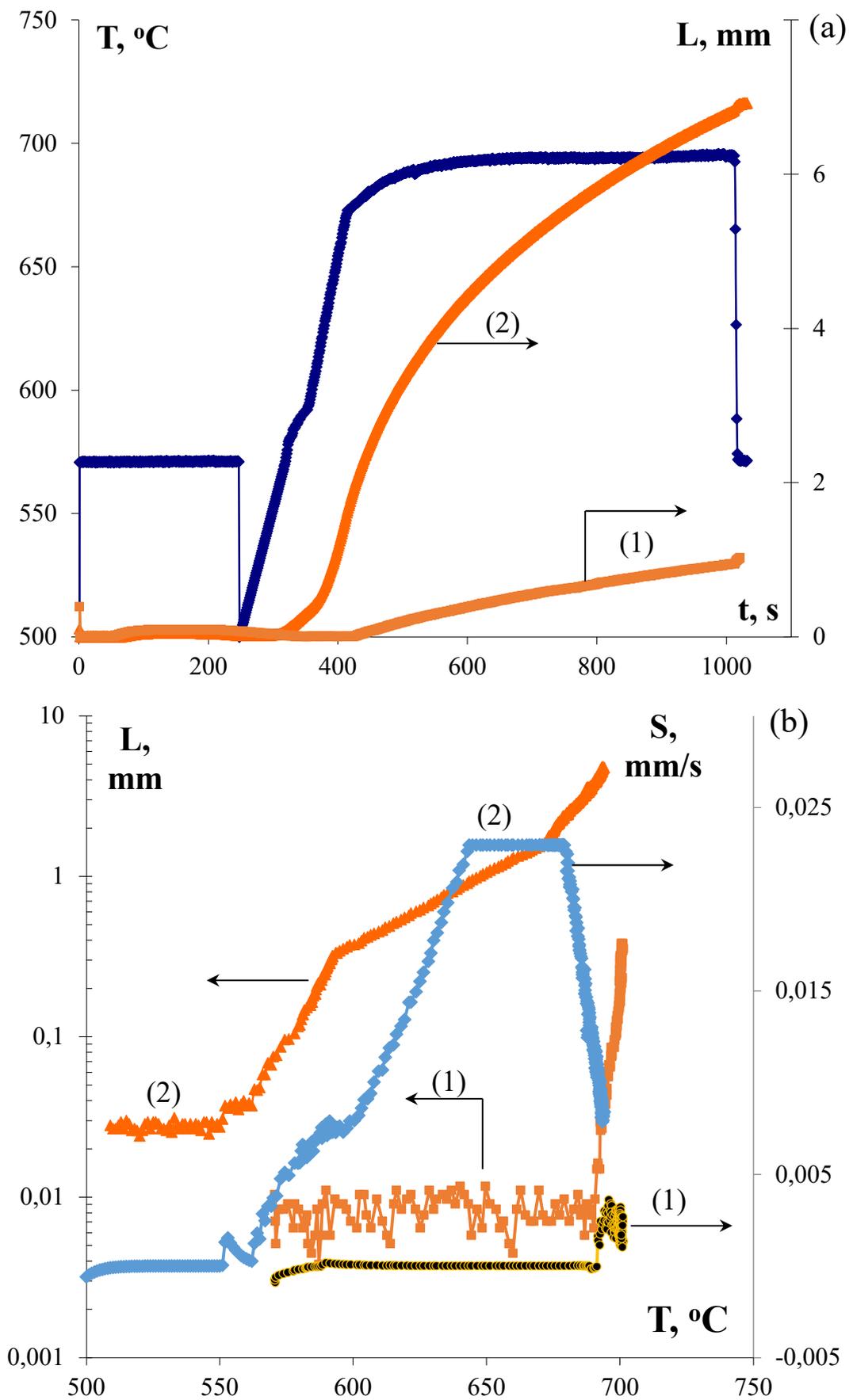

**Figure 9**

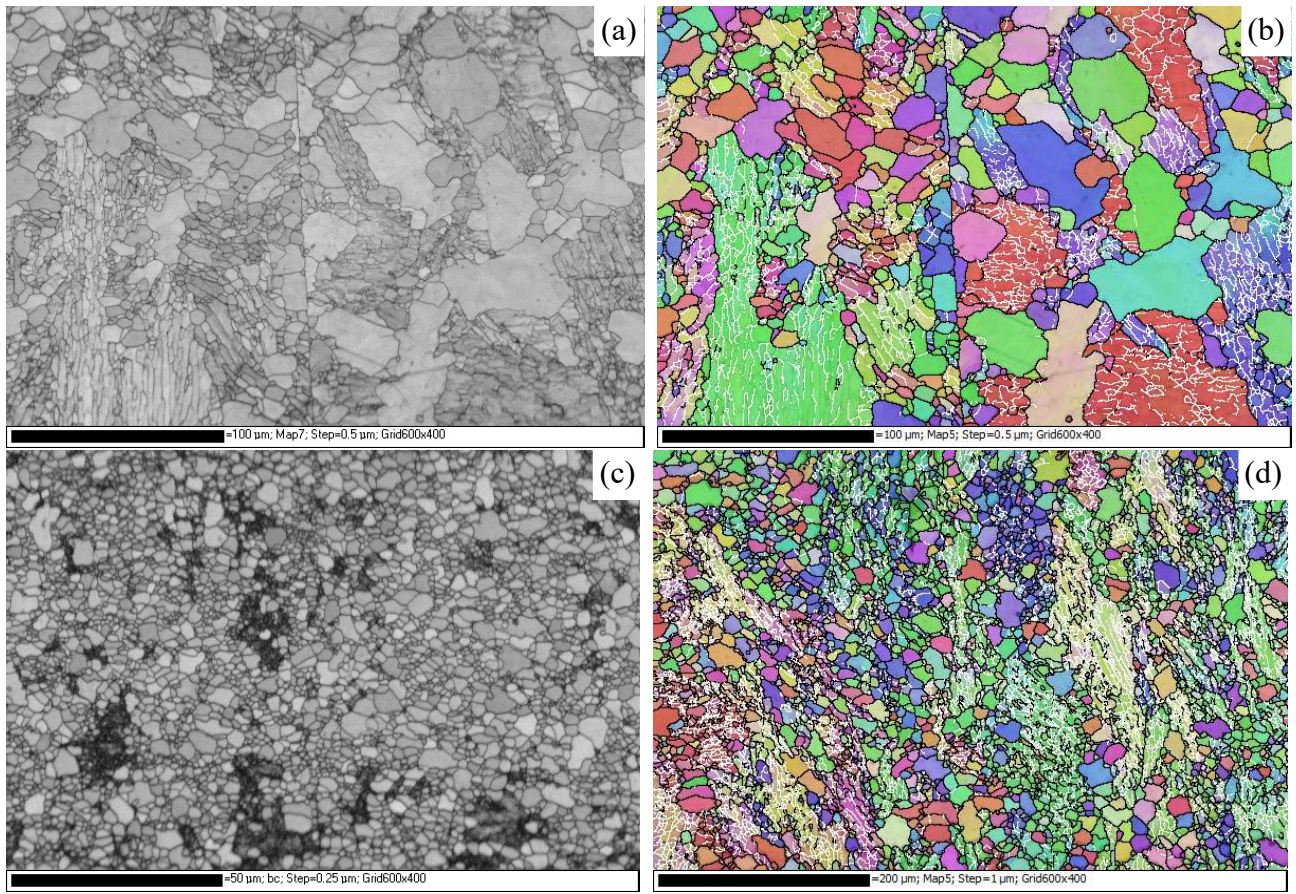

**Figure 10**

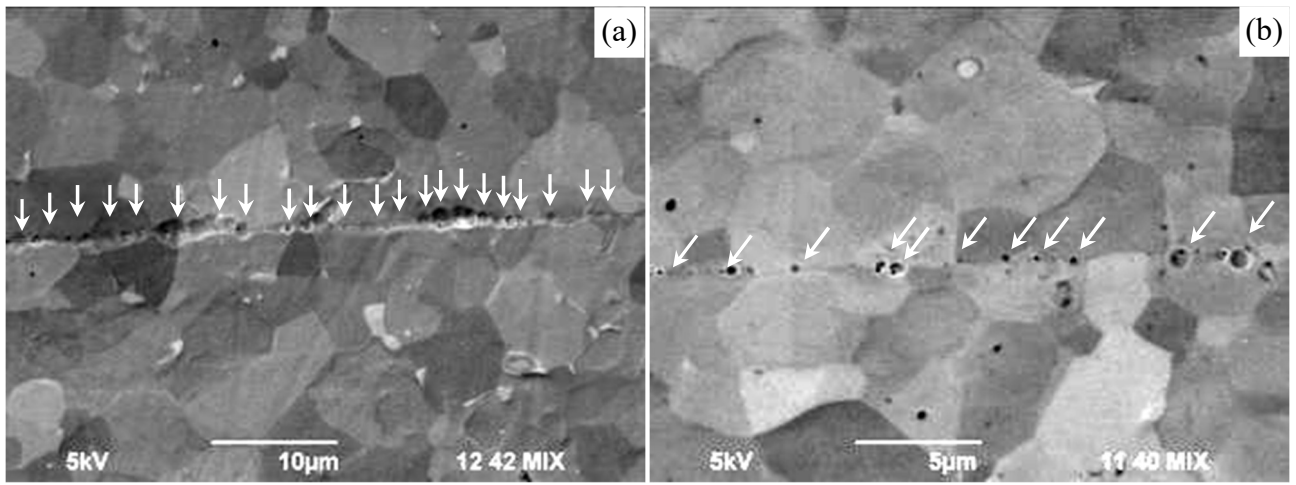

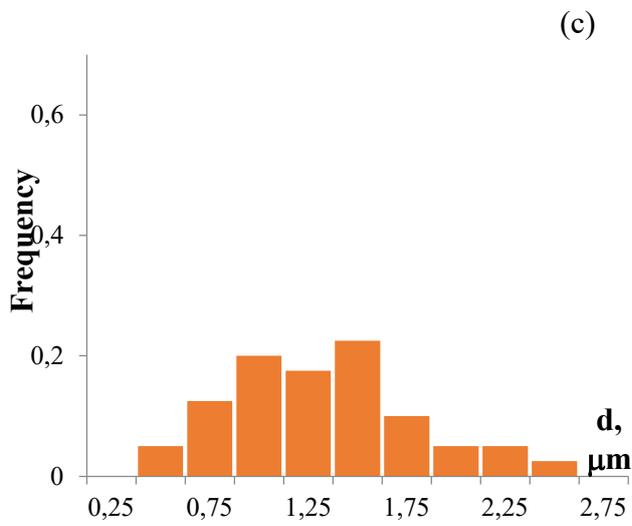
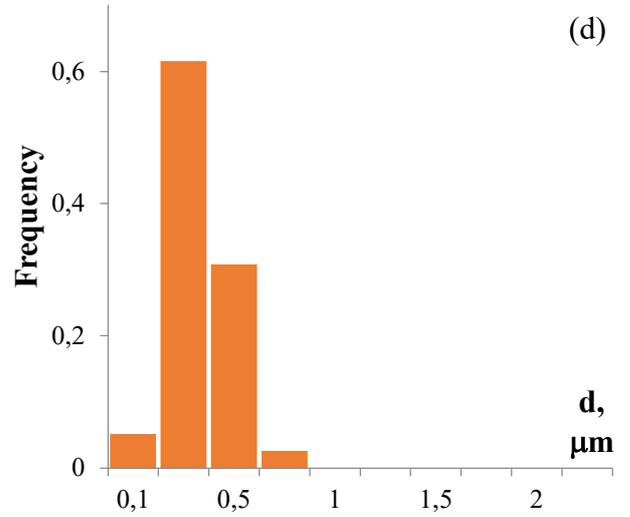

**Figure 11**

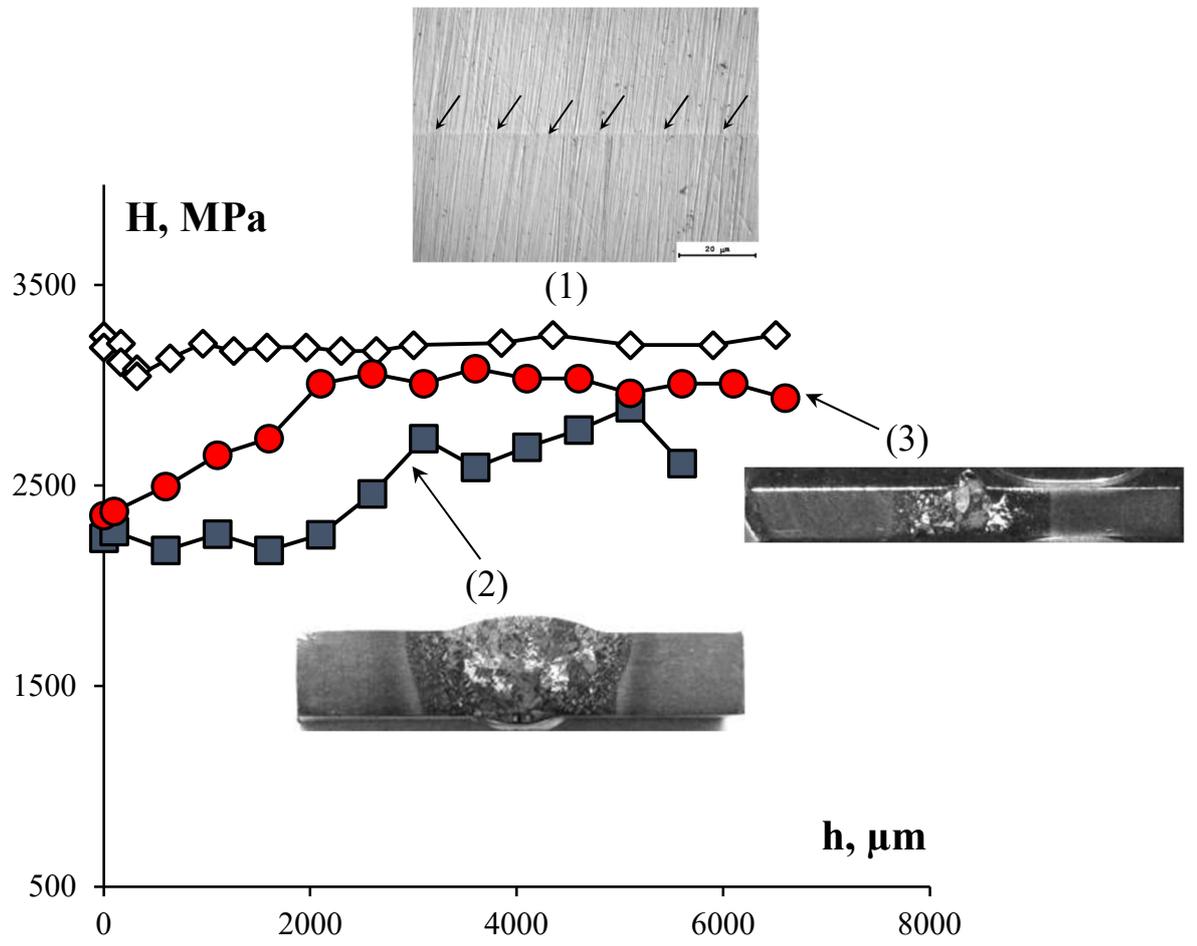

**Figure 12**

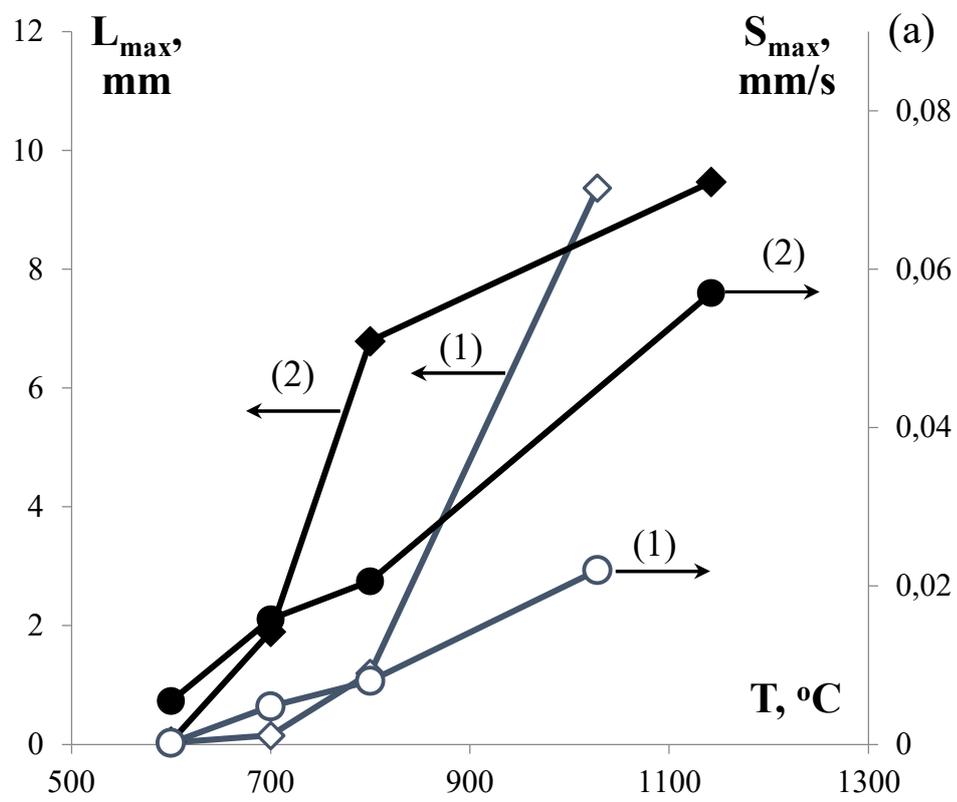

**Figure 13a**

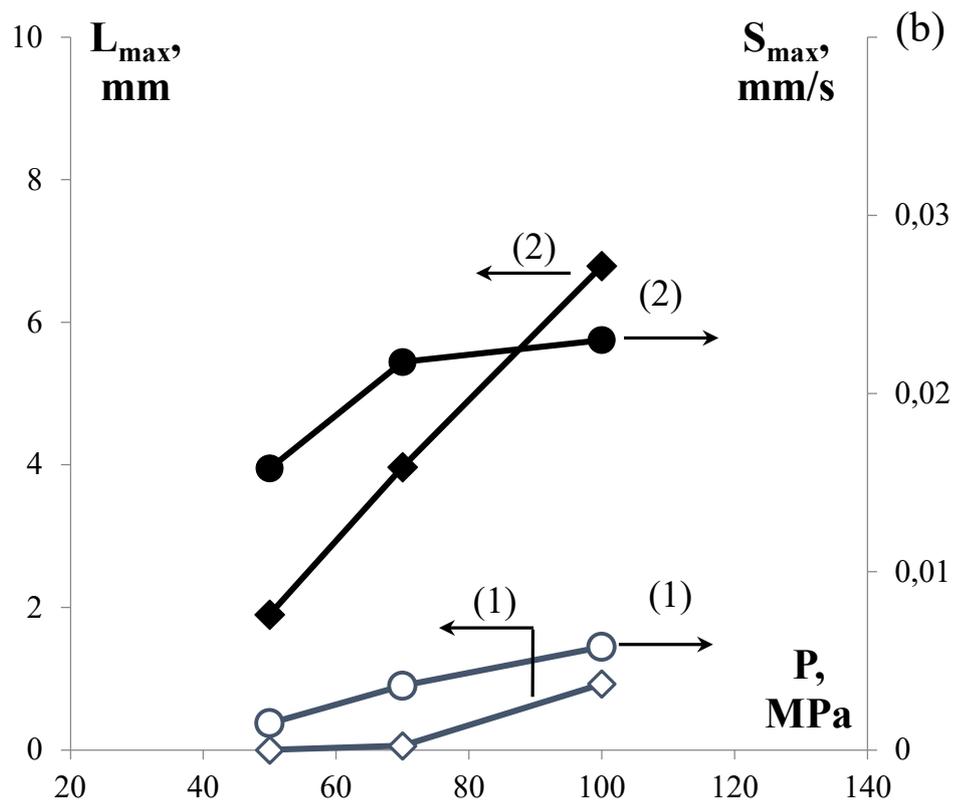

**Figure 13b**

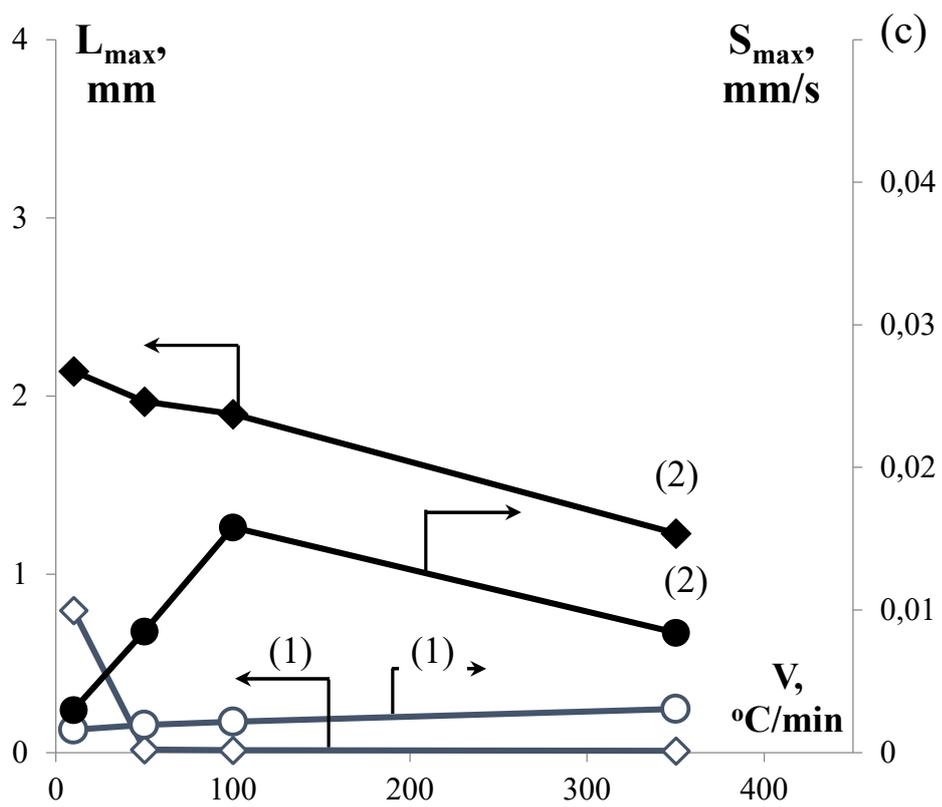

**Figure 13c**

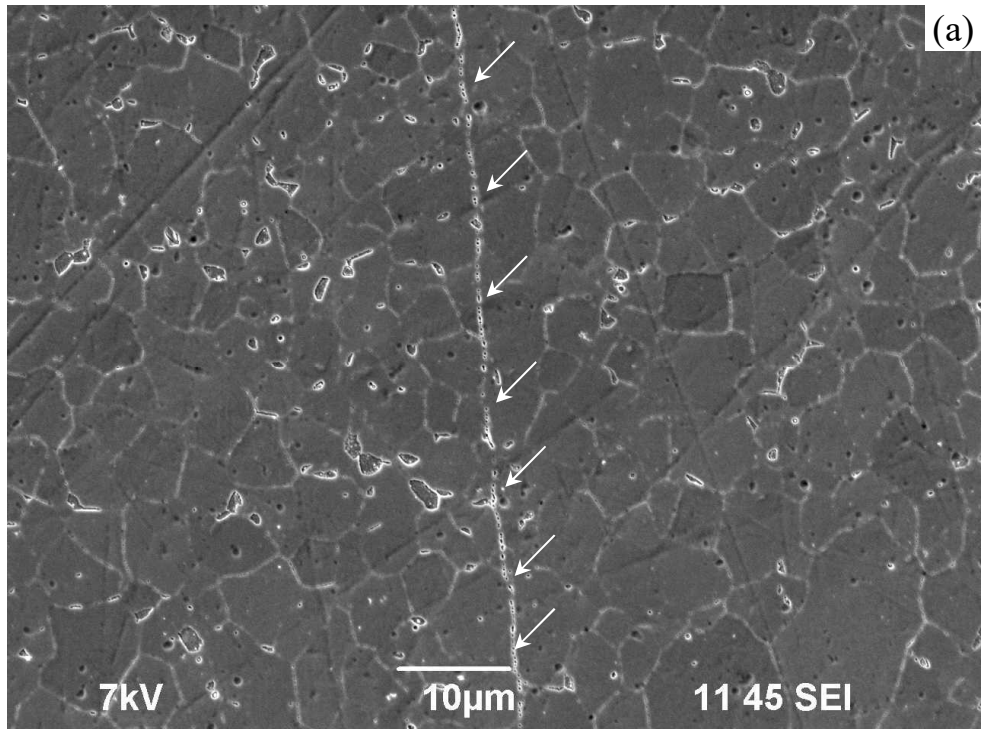

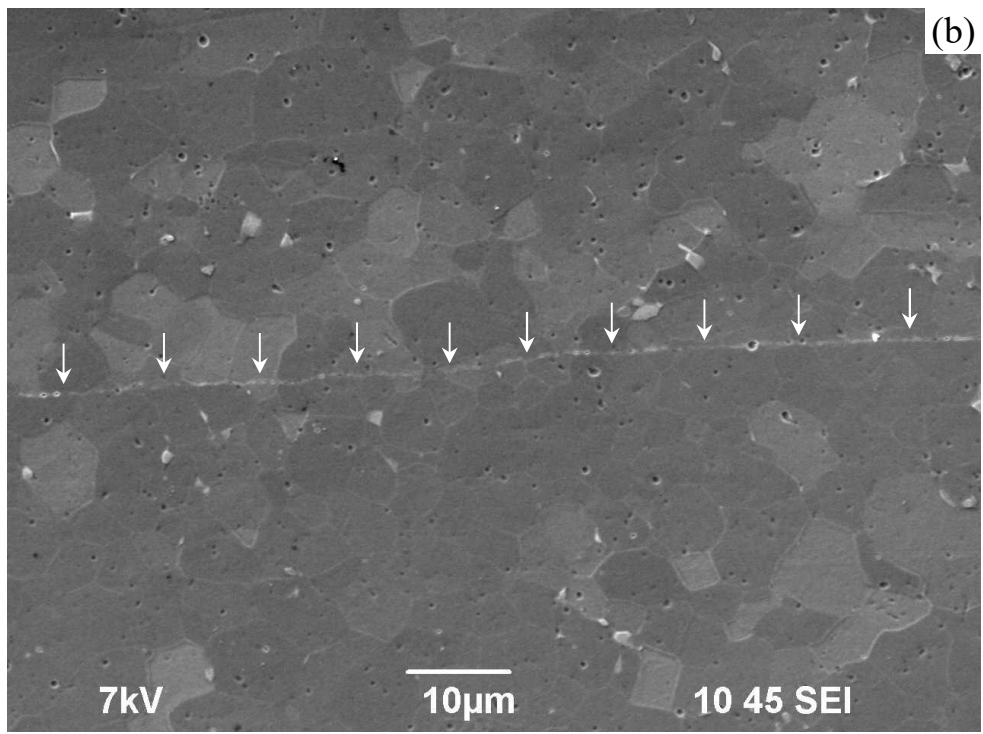

**Figure 14**

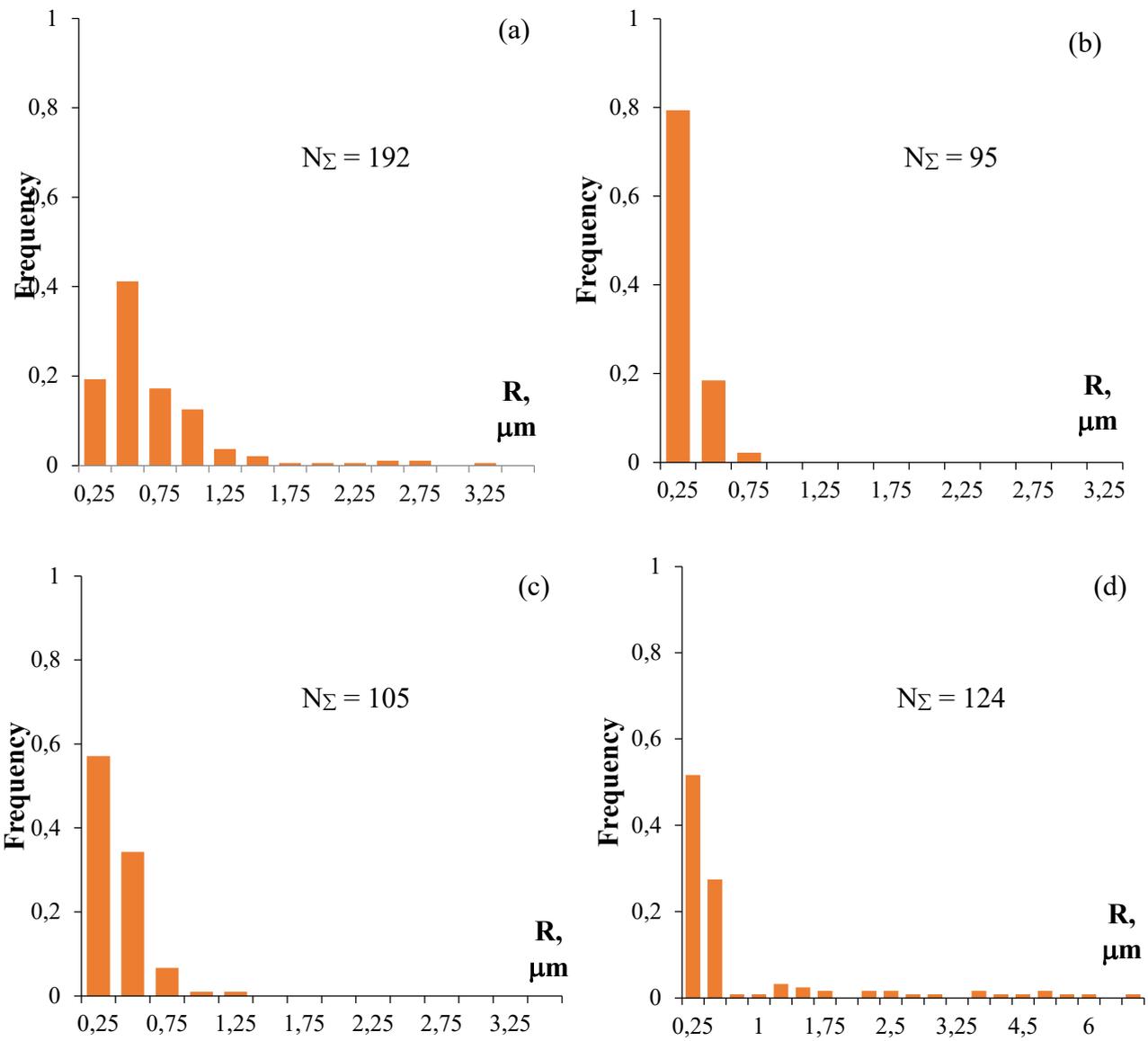

**Figure 15**

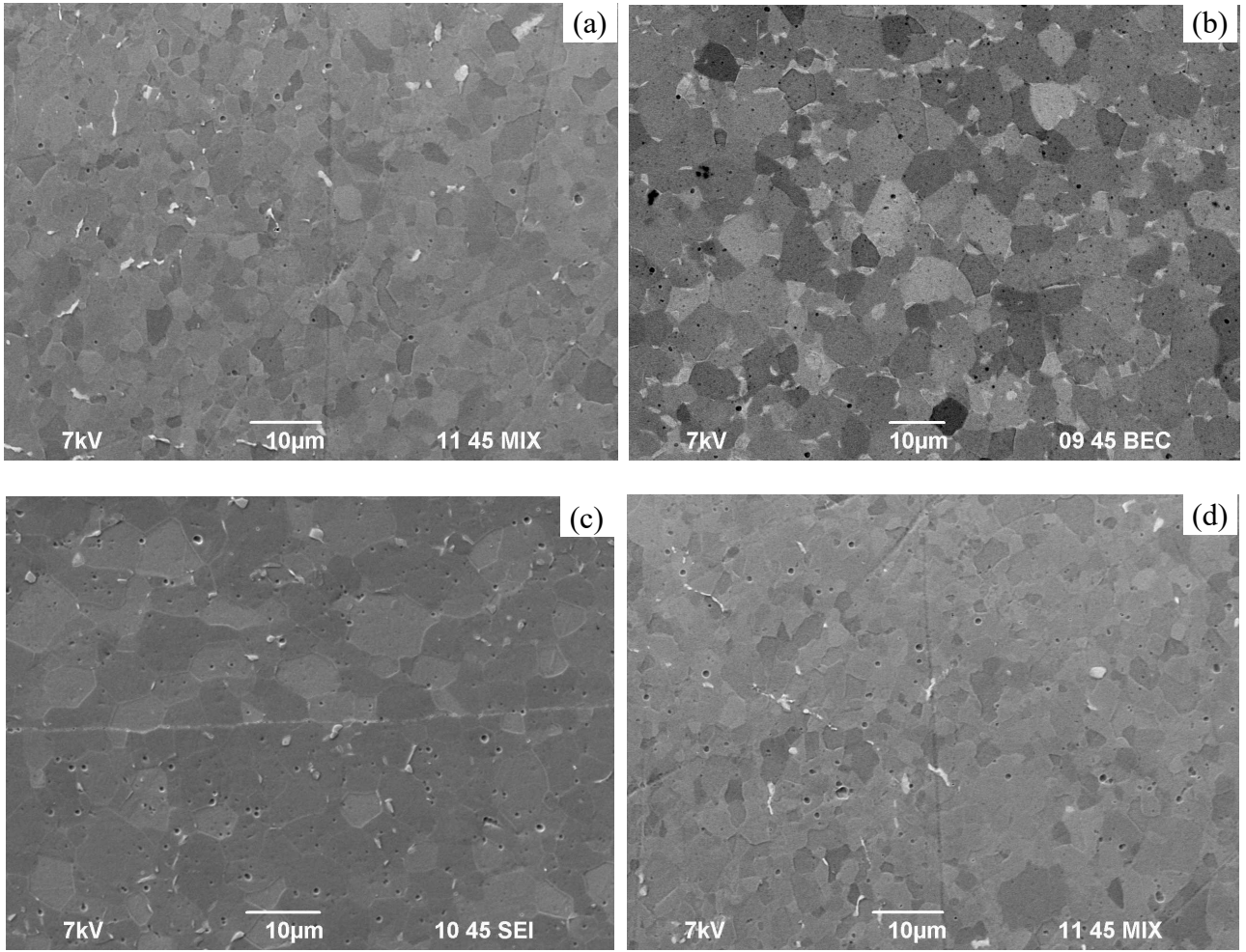

**Figure 16**

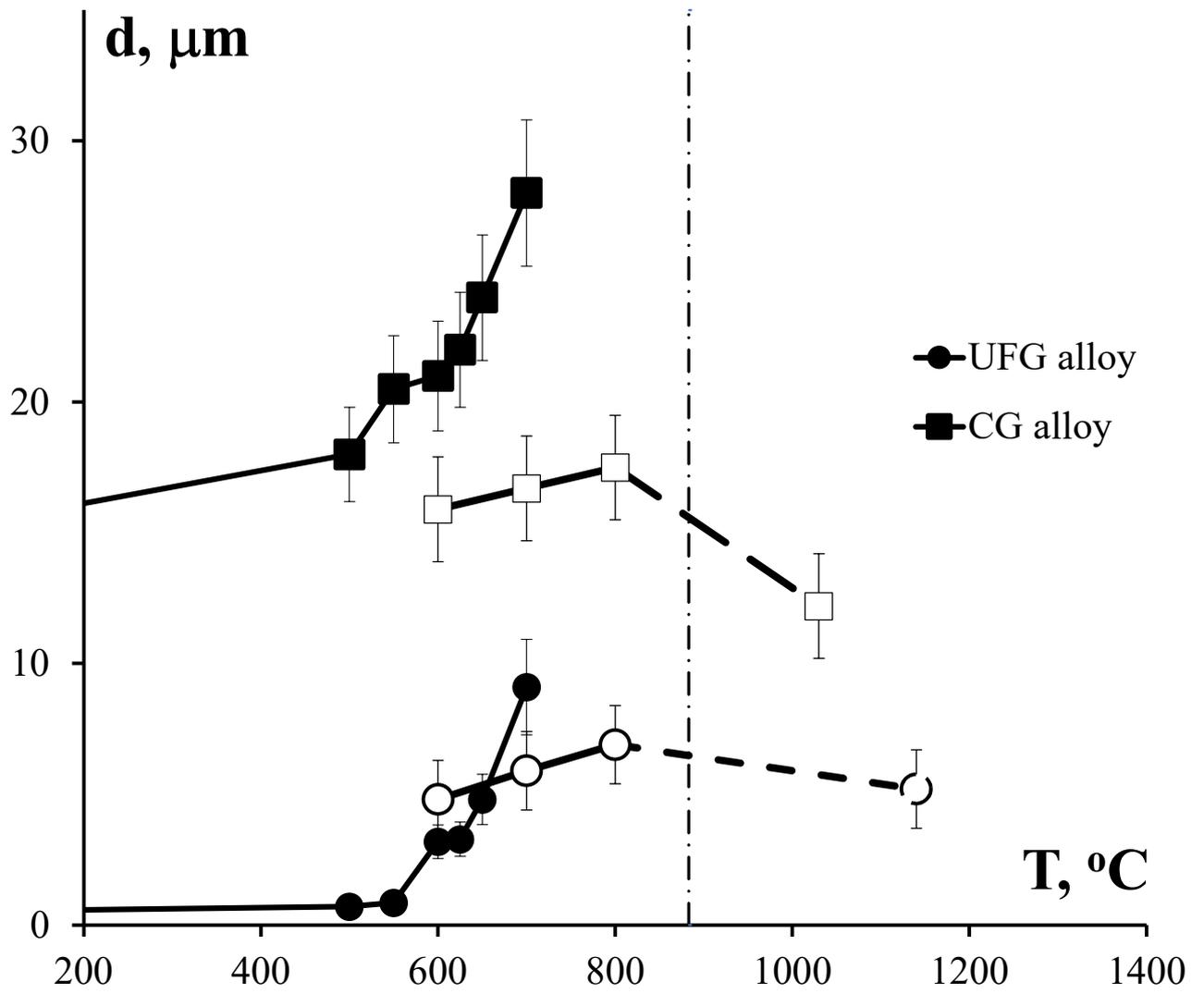

**Figure 17**

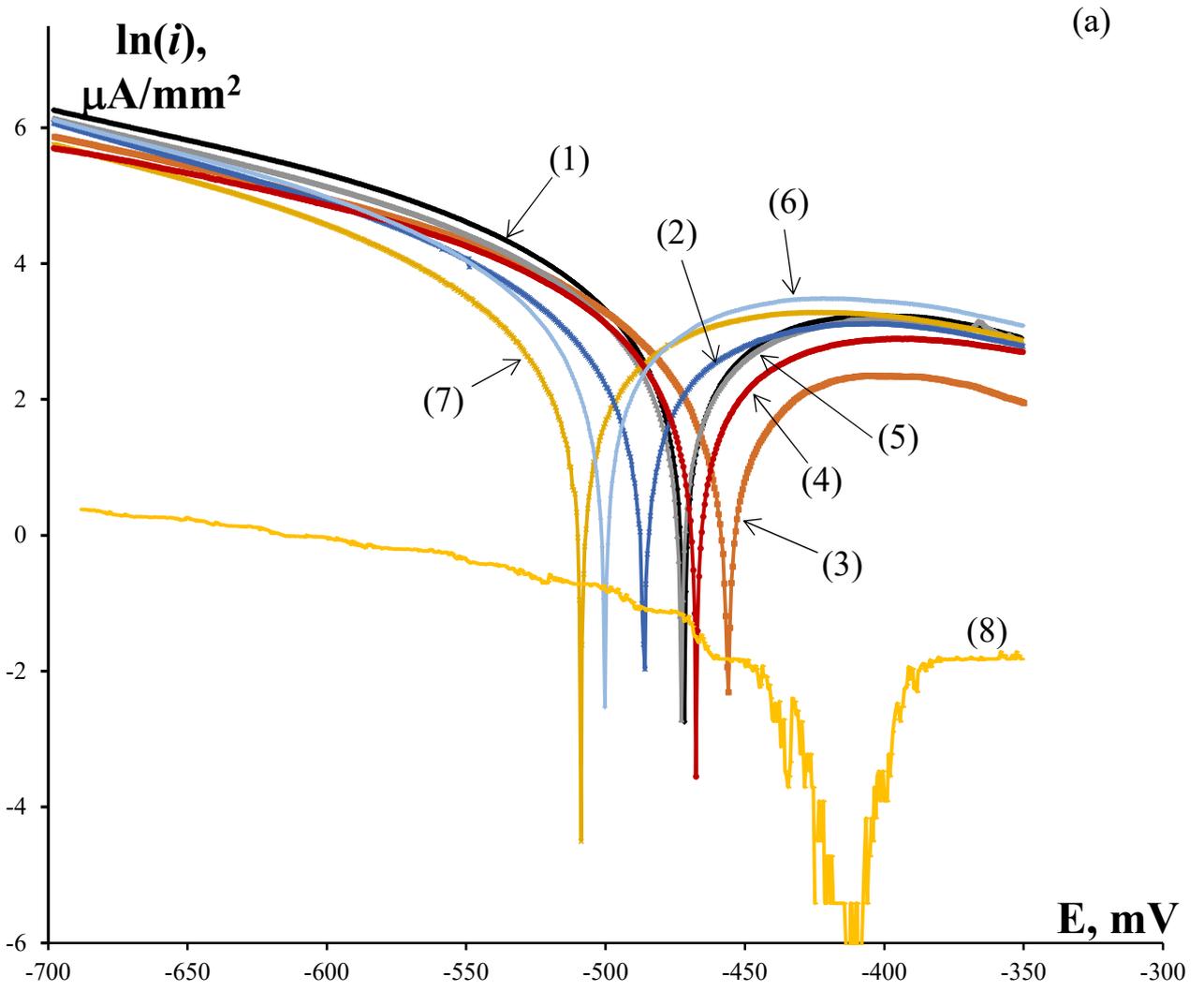

**Figure 18a**

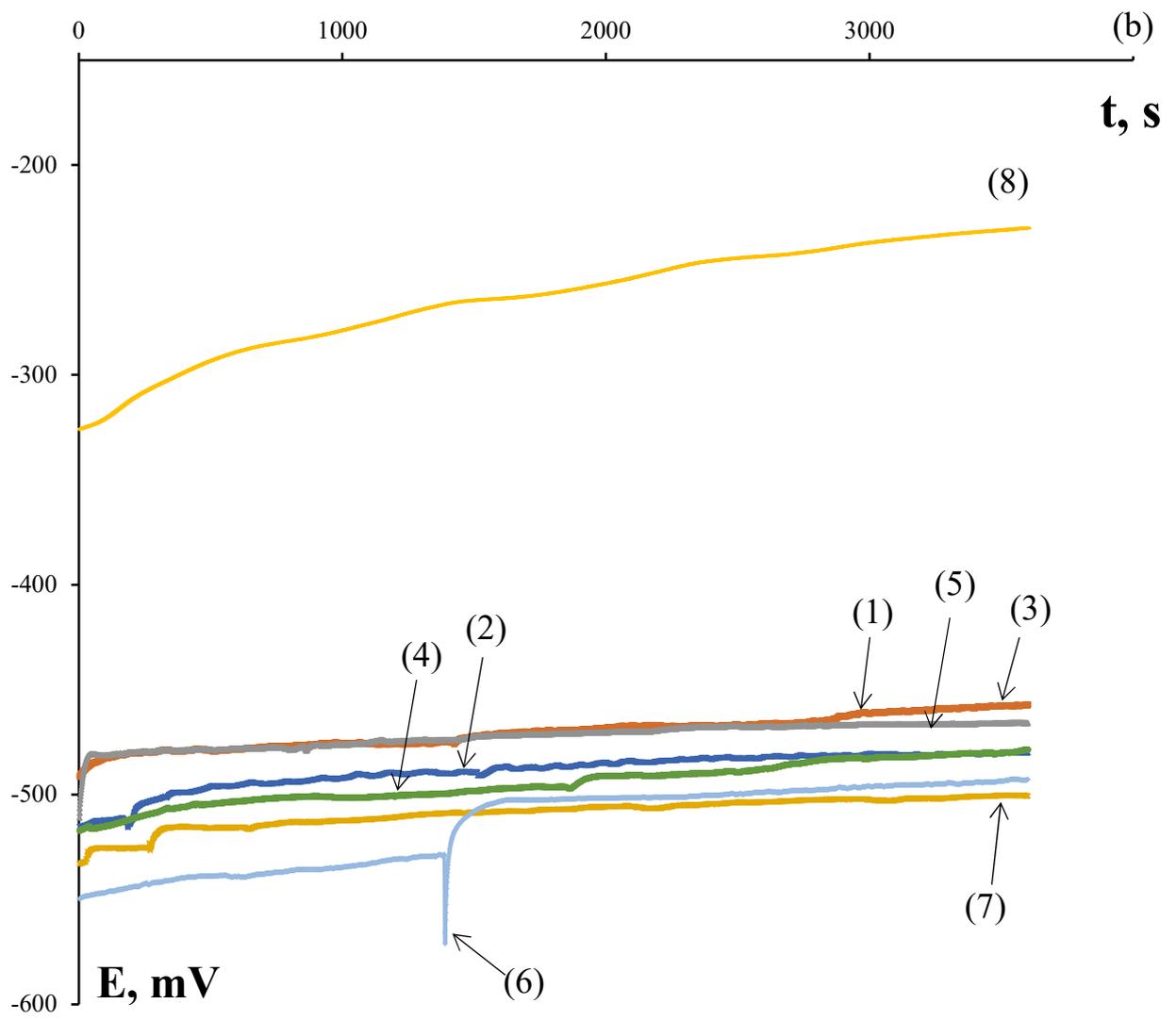

**Figure 18b**

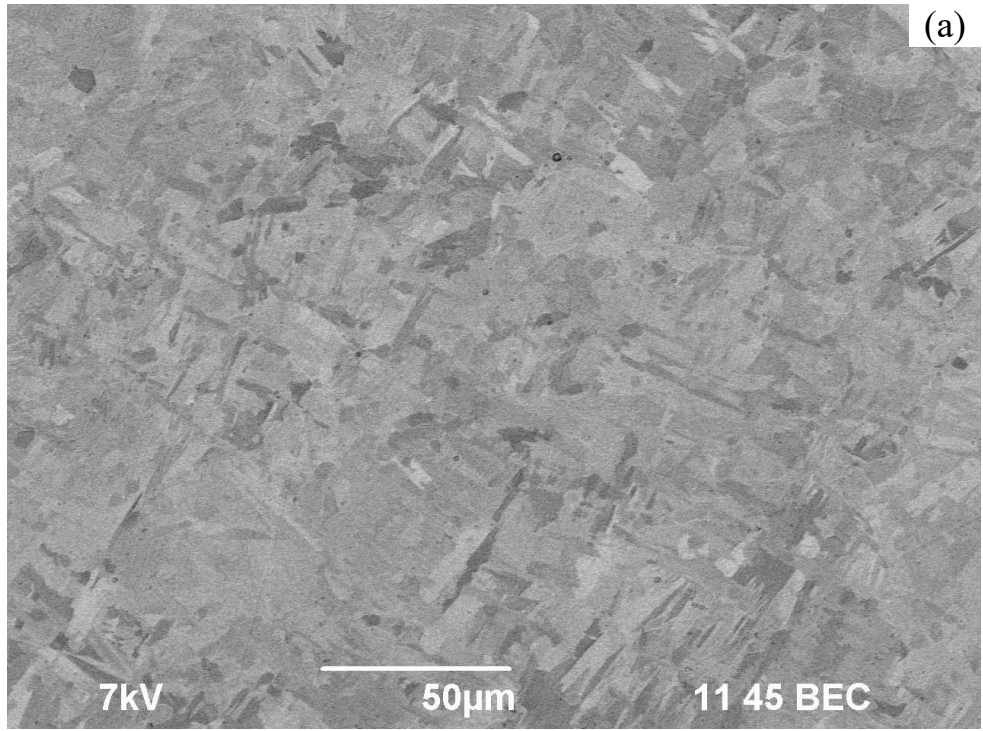

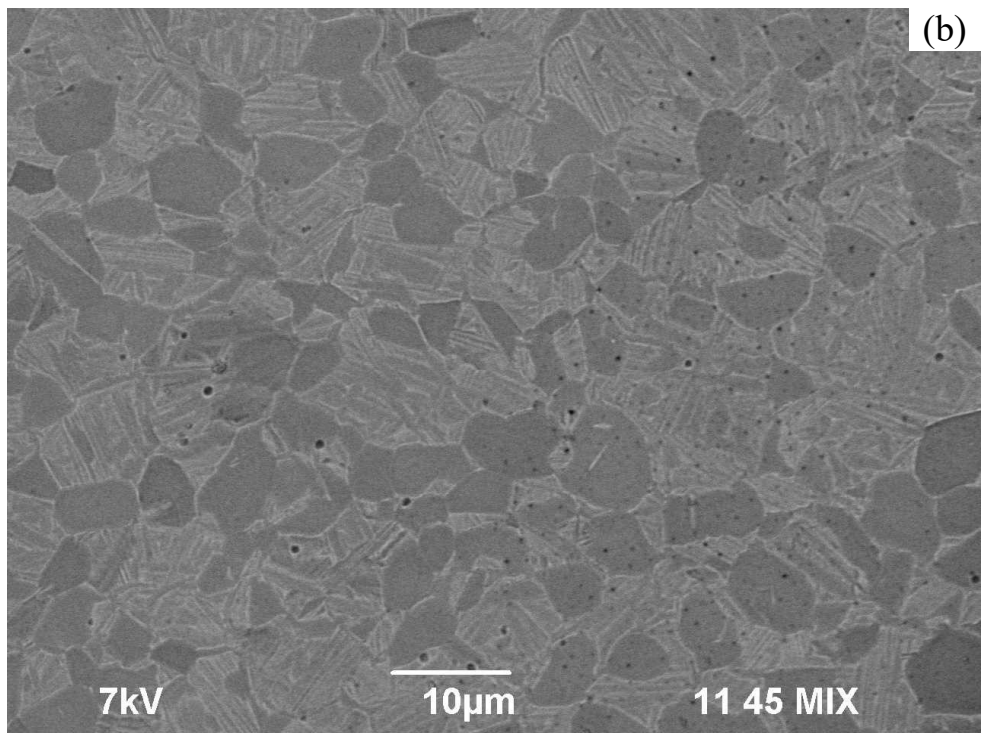

**Figure 19**

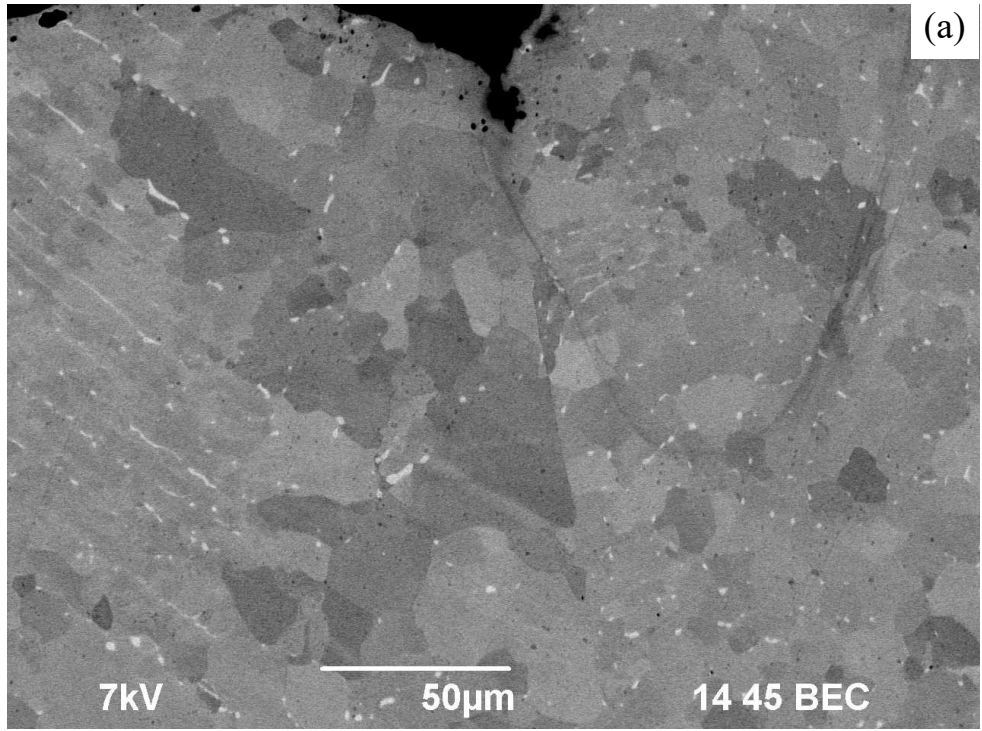
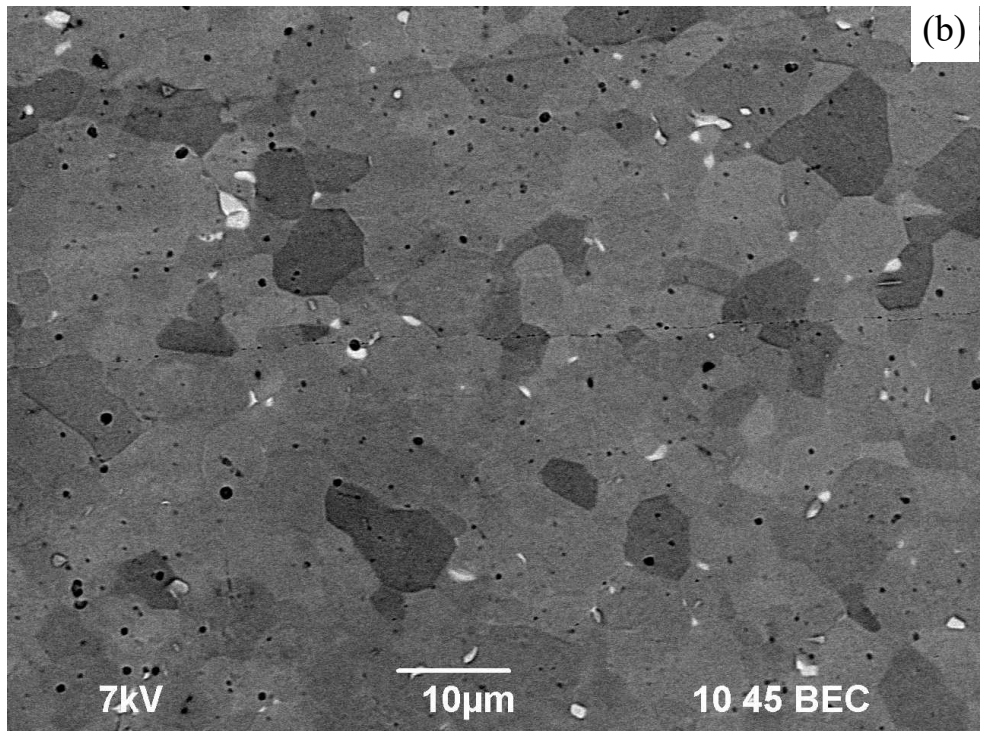

**Figure 20**

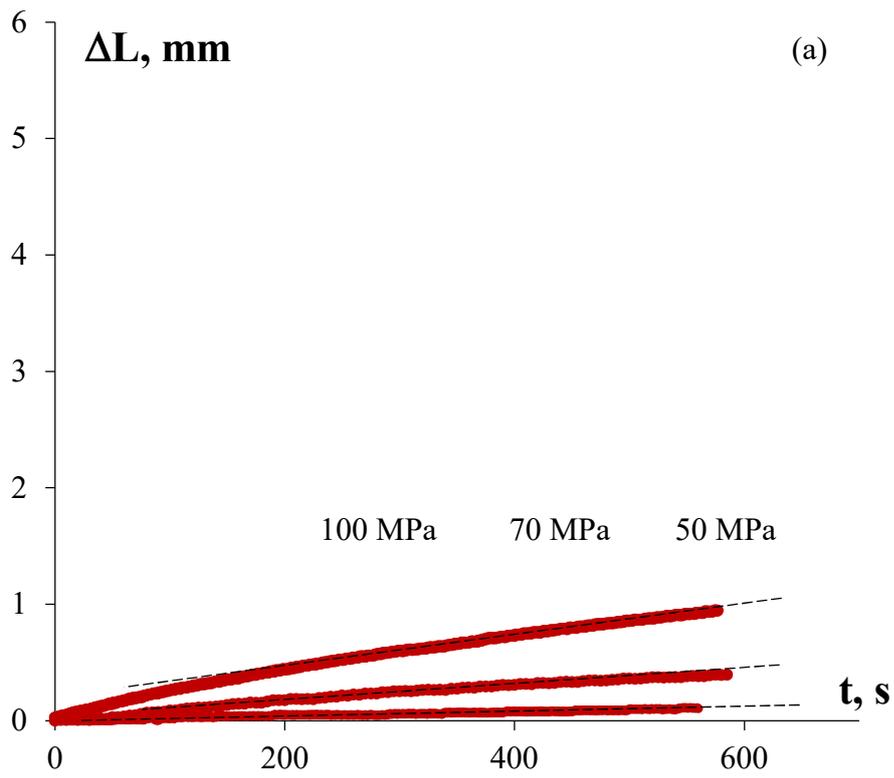

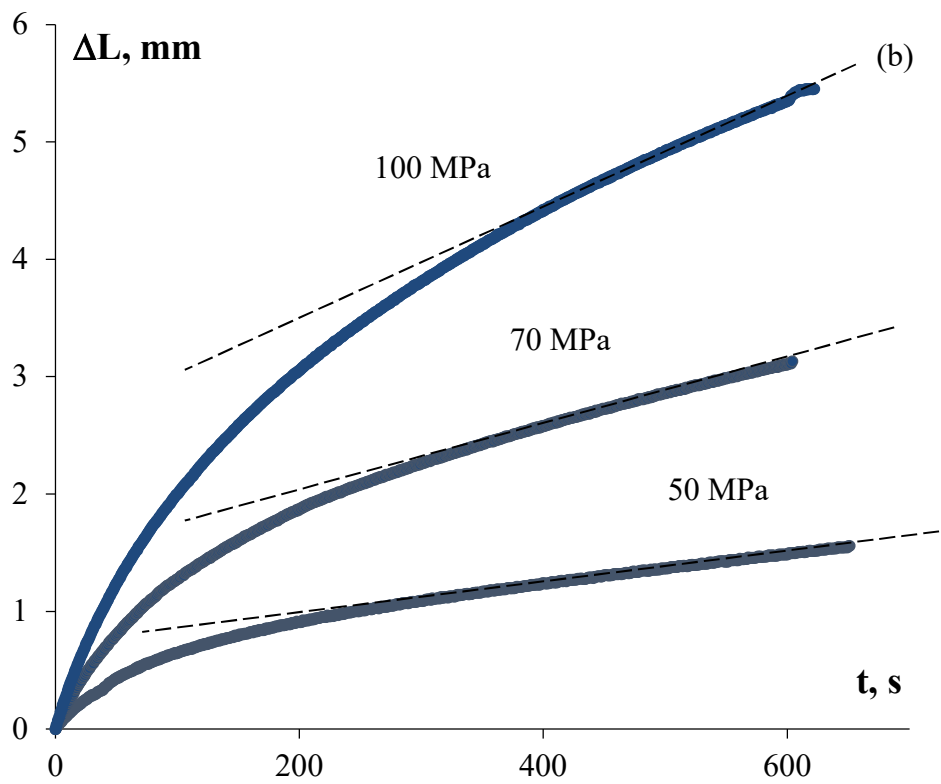

**Figure 21**

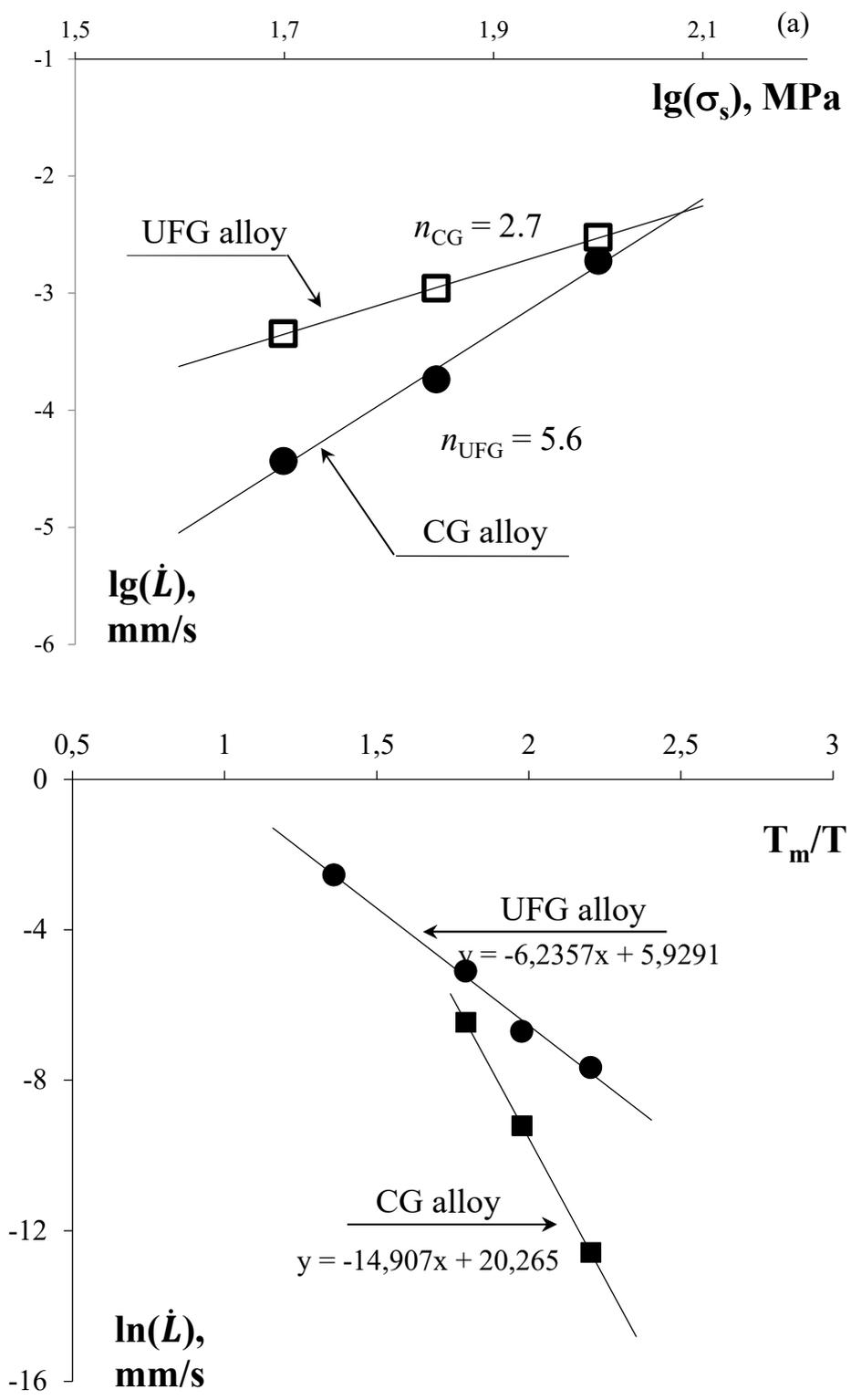

**Figure 22**

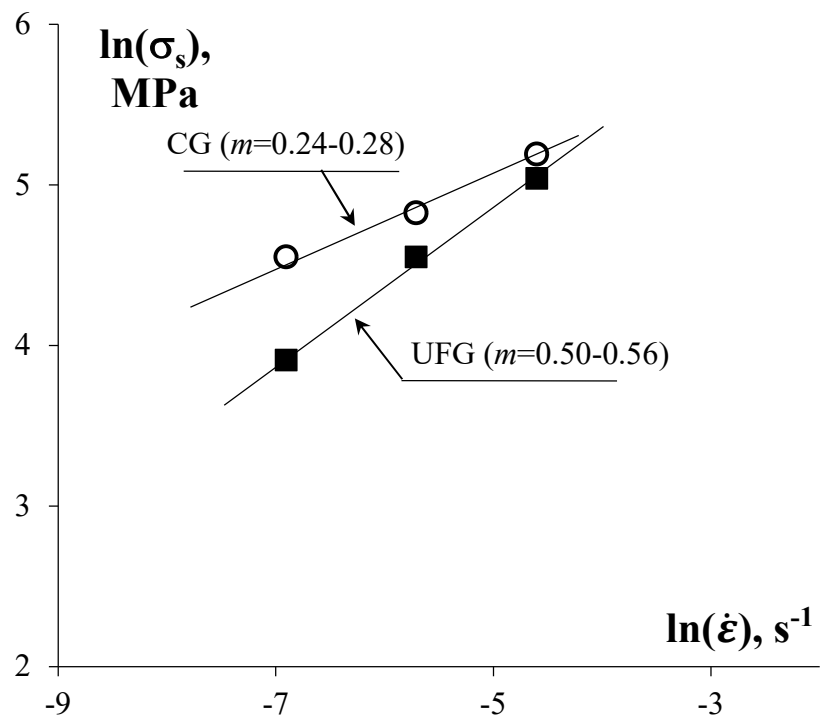

**Figure 23**

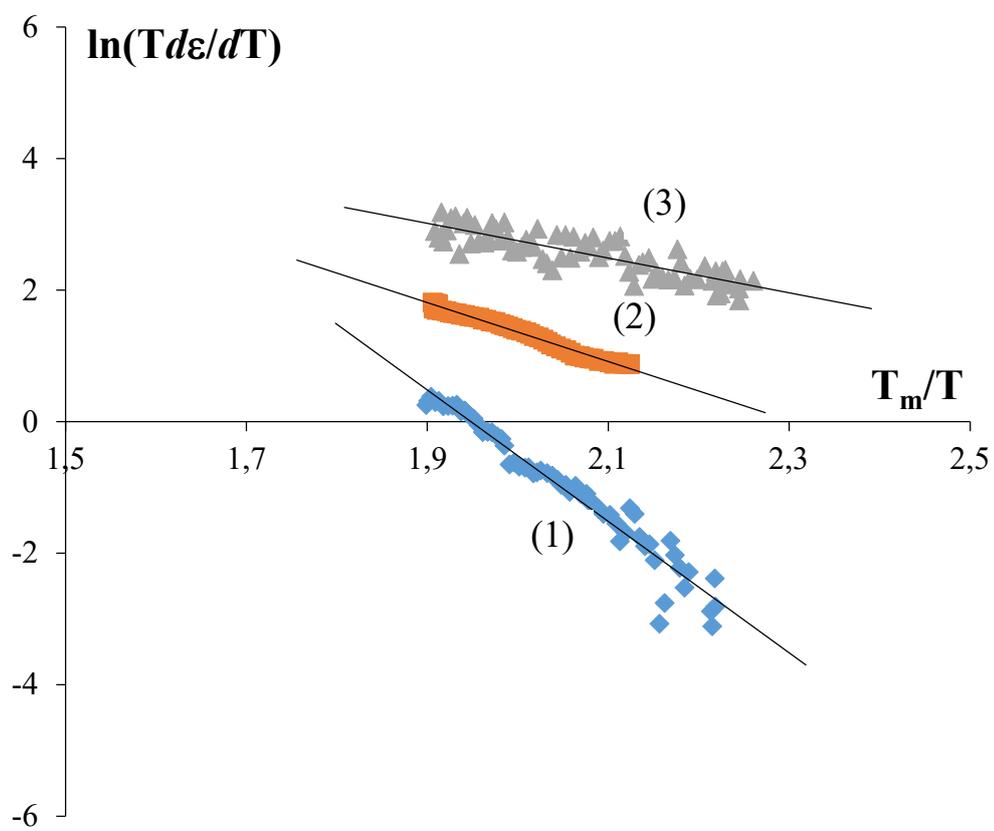

**Figure 24**